\documentclass{aastex7}
\usepackage{bm}
\usepackage{CJKutf8}
\usepackage{amsmath}
\usepackage{comment}

\pdfoutput=1

\newcommand{\imgvar}{x_i,y_j,t_k}
\newcommand{\imgvartb}{x_i,y_j,t_{k-1}}

\shorttitle{Multi-Scale Flare Ribbon Structure}
\shortauthors{Corchado-Albelo et al.}

\begin{document}

\title{Evolution of Spatial Complexity in Flare Ribbon Substructure and Its Relationship to Magnetic Reconnection Dynamics}

\correspondingauthor{Marcel F. Corchado-Albelo}
\author[0000-0003-1597-0184]{Marcel~F.~Corchado Albelo}
\email{marcel.corchado@colorado.edu}
\altaffiliation{DKIST Ambassador}
\affiliation{Department of Astrophysical and Planetary Sciences, University of Colorado Boulder, 2000 Colorado Avenue, Boulder, CO 80305, USA}
\affiliation{National Solar Observatory, 3665 Discovery Drive, Boulder, CO 80303, USA}
\affiliation{Laboratory for Atmospheric and Space Physics, University of Colorado Boulder, 3665 Discovery Drive, Boulder, CO 80303, USA}

\author[0000-0001-8975-7605]{Maria~D.~Kazachenko}
\email{maria.kazachenko@lasp.colorado.edu}
\affiliation{Department of Astrophysical and Planetary Sciences, University of Colorado Boulder, 2000 Colorado Avenue, Boulder, CO 80305, USA}
\affiliation{National Solar Observatory, 3665 Discovery Drive, Boulder, CO 80303, USA}
\affiliation{Laboratory for Atmospheric and Space Physics, University of Colorado Boulder, 3665 Discovery Drive, Boulder, CO 80303, USA}

\author[0000-0001-9726-0738]{Ryan~J.~French}
\email{ryan.french@lasp.colorado.edu}
\affiliation{Laboratory for Atmospheric and Space Physics, University of Colorado Boulder, 3665 Discovery Drive, Boulder, CO 80303, USA}

\author[0000-0002-5871-6605]{Vadim M. Uritsky}
\email{uritsky@cua.edu}
\affiliation{Catholic University of America 620 Michigan Ave., N.E. Washington, DC, 20064, USA}
\affiliation{NASA Goddard Space Flight Center 8800 Greenbelt Ave. Greenbelt, MD, 20771, USA}

\author[0000-0002-8767-7182]{Emily Mason}
\email{emason@predsci.com}
\affiliation{Predictive Science Inc., 9990 Mesa Rim Rd., Suite 170, San Diego, CA 92121, USA}

\author[0000-0002-3229-1848]{Cole A. Tamburri}
\email{cole.tamburri@colorado.edu}
\altaffiliation{DKIST Ambassador}
\affiliation{Department of Astrophysical and Planetary Sciences, University of Colorado Boulder, 2000 Colorado Avenue, Boulder, CO 80305, USA}
\affiliation{National Solar Observatory, 3665 Discovery Drive, Boulder, CO 80303, USA}
\affiliation{Laboratory for Atmospheric and Space Physics, University of Colorado Boulder, 3665 Discovery Drive, Boulder, CO 80303, USA}

\author[0000-0003-4065-0078]{Rahul Yadav}
\email{rahul.yadav@lasp.colorado.edu}
\affiliation{Laboratory for Atmospheric and Space Physics, University of Colorado Boulder, 3665 Discovery Drive, Boulder, CO 80303, USA}

\author[0000-0001-6886-855X]{Benjamin~J.~Lynch}
\email{bjlynch@epss.ucla.edu}
\affiliation{Space Sciences Laboratory, University of California, Berkeley, CA 94720, USA}
\affiliation{Department of Earth, Planetary, and Space Sciences, University of California, Los Angeles, CA 90095, USA}

\begin{abstract}
Recent three-dimensional flare models suggest that flare-ribbon substructure is linked to the fragmentation of the reconnecting current sheet in the corona. Flare-ribbon substructure can therefore potentially serve as a unique diagnostic tool {for} physical processes in the flare current sheet. {In this paper, we describe a new method to quantify the evolution of ribbon substructure, which first extracts the ribbon’s leading bright front and then quantifies its morphology using the box-counting dimension and Correlation Dimension Mapping (CDM).} We first test our method using synthetic observations. We then apply it to an M6.5-class solar flare on 2015 June 22 observed by the Interface Region Imaging Spectrograph (IRIS) 1330 \AA{} Slit-Jaw Imager (SJI). We find that when the flare ribbon boundary has {more} multi-spatial-scale features (higher box-counting dimension), hard X-ray (HXR) emission and magnetic-reconnection rates are the strongest. We also find that the flare-ribbon complexity characterized by CDM has moderate correlation with the IRIS Si IV 1402.77 \AA{}\ non-thermal velocity (in the {negative-polarity} ribbon) and reconnection flux rates (in ribbons of both magnetic polarities). We conclude that the build-up of the spatial complexity of the ribbons at multiple spatial scales can serve as an observational proxy for current-sheet fragmentation in the corona.   
\end{abstract}

%%%%%%%%%%%%%%%%%%%%%%%%%%%%%%%%%%%%%%%%%%%%%%%%

\section{Introduction}\label{sec:intr}
Solar flares are characterized by localized emission of light within the solar atmosphere, spanning the full electromagnetic spectrum from radio waves to $\gamma$-rays (see the review by \citealt{Benz2017FlareObservations}). They are understood to be associated with the release of free magnetic energy stored in twisted and sheared coronal magnetic fields via magnetic reconnection (see reviews by \citealt{Priest2002TheFlares,Hudson2011GlobalFlares,Shibata2011SolarProcesses}). In two dimensions (2D), the relationship between magnetic energy release and the transport of mass, heat, and non-thermal particles during the flare can be described by the CSHKP or the standard two-ribbon flare model \citep{Carmichael1964AFlares,Sturrock1966ModelFlares,Hirayama1974TheoreticalProminences,Kopp1976MagneticPhenomenon}. An extension to three dimensions (3D) includes a similar physical description and is better suited to explain flare observations \citep{Longcope2007Modeling2004,Aulanier2012TheDimensions,Aulanier2013TheDimensions,Janvier2013TheDimensions}. {Particles accelerated during the flare through bremsstrahlung can produce hard X-rays and $\gamma$-rays (\citealt{Brown1971TheBursts})}. 
In addition, heated plasma fronts traveling along the magnetic field lines can heat the chromosphere and enhance the emission at various wavelengths (e.g. H$_\alpha$ and ultraviolet; \citealt{Magara1996NumericalFlares,Huang2014HEruption}). The localized areas of enhanced emission (e.g. 1330 \AA{}, 1400 \AA{}, 1600 \AA{}, and 1700 \AA{}) in the upper chromosphere ($\approx$ 1000 to 2000 km above the solar surface) are called flare ribbons (e.g. \citealt{Forbes2000,Fletcher2001TheRibbons,Qiu2012HEATINGFUNCTIONS,Kazachenko2017}). The brightest and most compact sources of flare-ribbon emission are referred to as flare kernels \citep{Donnelly1971ExtremeResults, Kurokawa1988CloseFlare,Kitahara1990High-resolutionFlare, Warren2001UltravioletEmission,Alexander2006TemporalFlares,Nishizuka2009THEFLARE,Kowalski2017The29}. The morphology of flare ribbons is interpreted as a signature of heat and non-thermal particles precipitating from a coronal reconnecting source into the dense chromosphere at the footpoints of newly reconnected flare loops (\citealt{Forbes1984NumericalRegion,Forbes2000,Fletcher2001TheRibbons,Asai2004FlareRate,Qiu2012HEATINGFUNCTIONS,Longcope2014AFLARE,Graham2015TEMPORALFLARE,Li2017ImagingFlare,Priest2017Flux-RopeReconnection,Afanasyev2023HybridRegion,Fan2024A11158}). 

Some observational studies have focused on exploring the emission morphology of the so-called flare ribbon fine-structure or substructure to better understand the mechanisms through which energy is released and transported in the corona and chromosphere during a flare \citep{Barta2011SPONTANEOUSOBSERVATIONS,Karlicky2014SolarProcesses,Jing2016UnprecedentedTelescope,Parker2017ModelingShear,French2021ProbingDynamics,Kazachenko2022InvitedMagnetism,Pietrow2024SpectralRibbons,ThoenFaber2025High-resolutionStructures,Yadav2025Multi-lineViSP/DKIST,French2025EvolutionFlare}. In particular, observations of wave and swirl-like structures \citep{Brannon2015SPECTROSCOPICWAVES,Li2015,Parker2017ModelingShear}, ribbon front fragmentation \citep{Naus2022CorrelatedFlare,Cannon2023Magnetic22,Polito2023SolarSpectroscopy}, quasi-periodic pulsations \citep{Clarke2021Quasi-periodicFlare} and bursty flare emission \citep{Collier2024LocalisingFlare,Purkhart2025SpatiotemporalAcceleration,French2025DualFlareb} demonstrate that flare ribbons are spatial-temporal fragmented structures. {\cite{CorchadoAlbelo2024InferringRates} showed that the magnetic reconnection rates derived from flare ribbon observations are related to particle acceleration for dozens of flares of various classes. Furthermore, for two of the flares in their study flare ribbon spatial-substructure is observed co-temporally with enhancements in the reconnection flux rates.} 

The plasmoid (or tearing-mode) instability \citep{Furth1963Finite-resistivityPinch,Loureiro2007InstabilityChains,Uzdensky2010FastRegime,Huang2012DistributionReconnection,Comisso2017PlasmoidSheets} has been invoked as a theoretical explanation of the spatial-temporal fragmentation of flare ribbon emission in connection to magnetic reconnection dynamics, and has been applied to numerical experiments of solar eruptions \citep{Karpen2012THEFLARES,Lynch2016RECONNECTIONERUPTIONS}. In flares, the plasmoid instability arises when the flaring current sheet becomes unstable and tears, and subsequently forms plasmoids, flux ropes, or magnetic islands in 2D, within the current layer (\citealt{Shibata2001Plasmoid-induced-reconnectionReconnection}). In scenarios where multiple plasmoids are present in the current layer, they can merge or separate as they propagate toward the current sheet exhausts, the termination regions of the main current sheet, and trigger higher levels of plasmoid instability until their sizes approach the electron and ion kinetic scale of the current sheet plasma (\citealt{Kliem2000SolarReconnection,Shibata2001Plasmoid-induced-reconnectionReconnection}). This tearing of the current and merging of plasmoids allows magnetic reconnection to occur in bursts. In an attempt to link current observation capabilities to current sheet processes, recent magnetohydrodynamic (MHD) simulations of solar flares by \cite{Wyper2021IsSheet} and \cite{Dahlin2025DeterminingReconnection} have shown that current sheet plasmoids produced by the plasmoid instability map into the chromosphere as swirls and wave-like features along the flare ribbon fronts. In both studies, the appearance of ribbon features is more prevalent for the oblique tearing modes \citep{Daughton2011RolePlasmas}, which form plasmoids at an oblique angle to the guide field.

Although there are no direct magnetic measurements of coronal plasmoids, many observations are interpreted as plasmoid signatures. In the chromospheric and corona, bright blob-like features have been interpreted as evidence of current sheet plasmoids \citep{Ohyama1998XRayEjection, Kliem2000SolarReconnection, Takasao2012SIMULTANEOUSFLARE,Hayes2019,Lu2022ObservationalFlare,Kumar2025X-Ray/RadioSheets}. Furthermore, new adaptive optics systems on the Goode Solar Telescope (GST; \citealt{Goode2010TheEST,Schmidt2018WavefrontProminences}) have resolved the dynamical evolution of one such coronal structure in a post-reconnection magnetic loop system \citep{Schmidt2025ObservationsOptics}.  

Plasmoid instability has been also reported in the context of magnetic reconnection in different plasma environments. It has been identified as a candidate to produce substructures small-scale spiral features in Earth's aurora \citep{Huang2022AuroralRegion,Nakamura2025OutstandingReconnection}. {In the astrophysical context, plasmoid instability has been linked to substructure observations in numerical experiments of pulsar winds \citep{Cerutti2021FormationNebula}, accretion disks of black holes \citep{Ripperda2022BlackReconnection} and multiphase interstellar wind \citep{Fielding2023PlasmoidMedium}. Plasmoids have also been observed in magnetic reconnection laboratory experiments with collisionless \citep{Olson2016ExperimentalReconnection} and collisional-collisionless \citep{Zhao2022LaboratoryRegime} plasma regimes.} Therefore, the study of flare ribbon morphology substructure in the context of reconnection events should provide new perspectives into coronal current sheet reconnection physics. 

{In two-ribbon solar flares the coronal magnetic reconnection dynamics can be approximated by combining measurements of the photospheric magnetic field and flare ribbon morphology evolution.} \cite{Forbes2000WhatEjections} described the breakdown of the \textit{frozen flux} condition (\citealt{ALFVEN1942ExistenceWaves}), consistent with the general theory of 3D magnetic reconnection \citep{Hesse1988AReconnection,Schindler1988GeneralHelicity,Hesse2005OnCorona,Wyper2015QuantifyingLayers}, as
\begin{align}
    \frac{d\Phi}{dt} &= -\int\int_A(\nabla \times \textbf{R}^{\text{cor}})_{\parallel} dA  =\oint_{\mathcal{M}(\mathcal{B}(t))} \text{B}_{n}^{\text{phot}} \frac{d{\mathcal{B}}}{dt} dl \label{eq:flux}. 
 \end{align} 
{Here $d\Phi/dt$ represents the reconnection flux rate, $(\nabla \times \textbf{R}^\text{cor})_{\parallel}$ the curl of the non-ideal coronal electric-field ($\textbf{R}^\text{cor}$) parallel to the local magnetic field vector, $\mathcal{M}(\mathcal{B}(t))$ is the cumulative ribbon area boundary, $d{\mathcal{B}}/dt$ the multi-spatial-scale flare ribbon morphology evolution rate, and B$_{n}^{\text{phot}}$ the magnetic field normal to the solar surface. The reconnection flux rate ($d\Phi/dt$) represents the magnetic flux that changed connectivity due to the coronal magnetic reconnection process. This break in the \textit{frozen flux} condition is described by $(\nabla \times \textbf{R}^\text{cor})_{\parallel}$. The non-ideal coronal electric field is represented by the relevant terms of the generalized Ohm's law for coronal plasma (\citealt{Somov2006ThePlasma}). In the coronal context, Ohm's law is often approximated by its resistive form, such that the non-ideal electric field  is written as $\textbf{R}^\text{cor} \approx \eta\textbf{j}^\text{cor}$, where $\eta$ denotes the plasma resistivity. However, this description does not require explicit knowledge of the physical origin of the non-ideal electric field. Instead, the coronal reconnection process can be approximated using lower atmospheric observations, as described on the right-hand side of Equation~\ref{eq:flux}.} 

In this paper, we explore the relationship between the evolution of flare ribbon substructure and other indicators of coronal reconnection dynamics (e.g. ribbon-derived magnetic reconnection rates, HXR and ultraviolet (UV) emission, and non-thermal velocities derived from chromospheric and transition region emission). Additionally, we present a new methodology to characterize the spatial complexity of the flare ribbon based on a combination of the box-counting dimension and the generalized version of the correlation dimension mapping algorithm \citep{Mason2022StatisticalBoundary}. The paper is organized as follows. In Section \ref{sec:data_met} we describe the data used in this study and the new methodology to track the flare ribbons' bright leading edge and characterize their spatial complexity. In Section \ref{sec:res} we present our results from tracking the spatial and temporal evolution of flare ribbon substructure complexity for an M6.5 solar flare, and the comparison to other indicators of magnetic reconnection dynamics. In Section \ref{sec:dis} we discuss and interpret our results in the context of recent works in the field. Finally, we summarize our findings in Section \ref{sec:con}.
%%%%%%%%%%%%%%%%%%%%%%%%%%%%%%%%%%%%%%%%%%%%%%
\section{Data \& Methods}\label{sec:data_met}
\subsection{Event \& Data Description}\label{subsec:data}
\begin{figure}
    \centering
    \includegraphics[height=0.72\textheight,trim={15cm 17cm 15cm 18cm}]{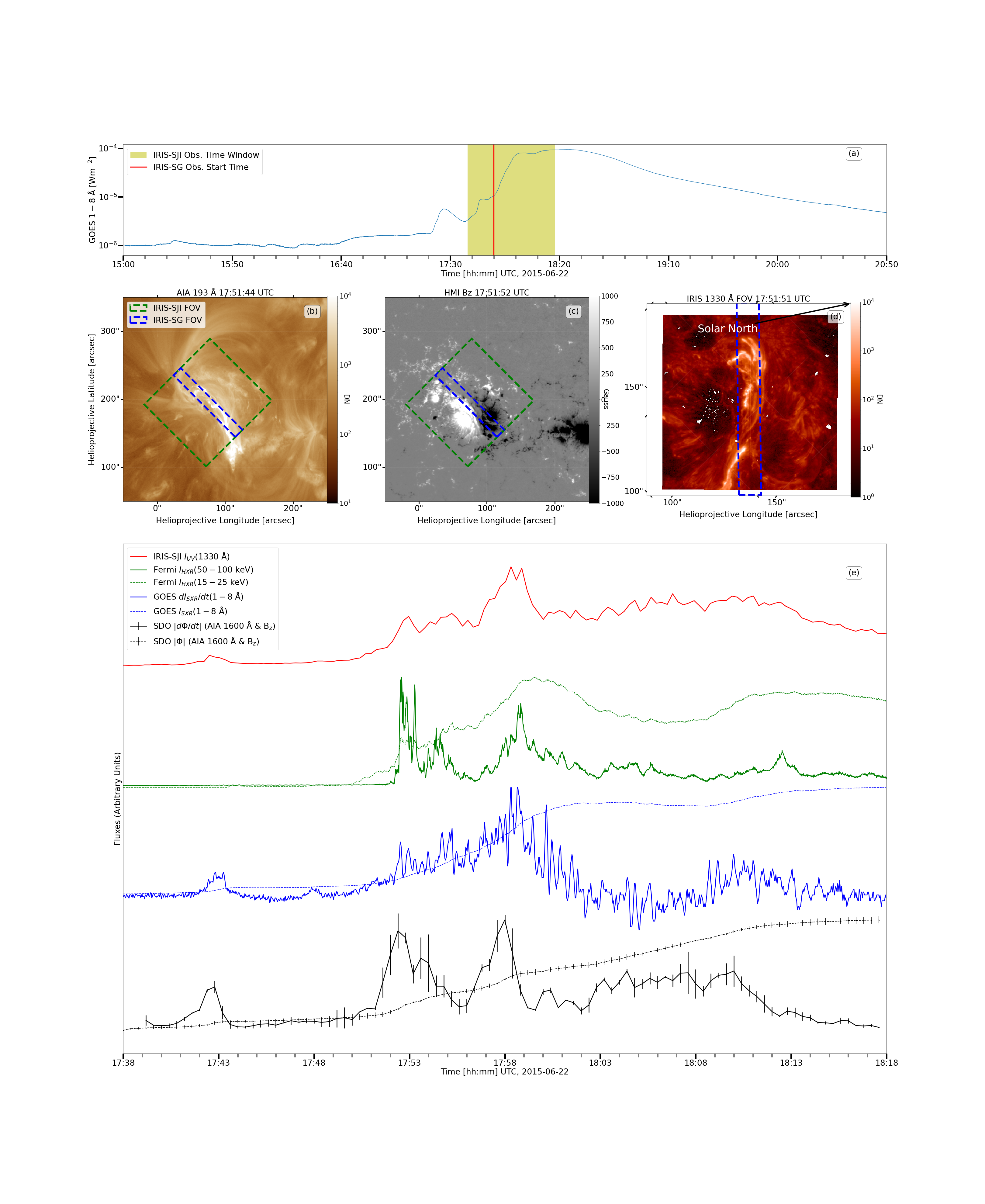}
    \caption{{Overview of the M6.5 flare in AR 12371 on 2015 June 22. (a) GOES $1-8$ \AA{} light curve; the IRIS-SJI analysis window is shaded yellow and the IRIS-SG start time is marked by the vertical line. (b) AIA $193$ \AA{} image at 17:51 UT, 
    (c) corresponding HMI B$_z$ magnetogram at 17:51 UT; (d) IRIS-SJI 1330 \AA{} at 17:51 UT. Green and blue boxes show the SJI and SG FOVs, respectively. 
    (e) Normalized time series of variables integrated over respective field of views: 
    IRIS 1330~\AA\ intensity ($I_{\rm UV}$; red), GOES 1--8~\AA\ ($I_{\rm SXR}$; blue dashed) and its time derivative ($dI_{\text{SXR}}/dt$; blue solid), Fermi ($I_{\rm HXR}$) 15--25~keV (green dashed) and 50--100~keV (green solid), and ribbon-derived reconnection flux ($\Phi$; black dashed) and rate ($d\Phi/dt$; black solid).
    }} 
    \label{fig:flare_cont}
\end{figure}

In this section we describe observations from the {Interface Region Imaging Spectrograph} (IRIS; \citealt{DePontieu2014TheIRIS,DePontieu2021AIRIS}), {Helioseismic Magnetic Imager} (HMI; \citealt{Hoeksema2014ThePerformance}) and {Atmospheric Image Assembly} (AIA; \citealt{Lemen2012TheSDO}) onboard the {Solar Dynamics Observatory} (SDO; \citealt{Pesnell2012TheSDO}), {Fermi Gamma-ray Space Telescope} (Fermi; \citealt{Atwood2009TheMission}), and {Geostationary Operational Environmental Satellite} (GOES; \citealt{Bornmann1996GOESDisturbances, Chamberlin2009NextSeries}) that we use to analyze an eruptive M6.5 flare on 2015 June 22. From the GOES soft X-ray (SXR) lightcurve the flare starts at  17:39 UT in National Oceanic and Atmospheric Administration (NOAA) Active Region (AR) 12371 close to the solar disk center. The magnetic configuration of AR 12371 at the time of the flare is shown in the middle panel of the second row of Figure~\ref{fig:flare_cont}. Additionally, this figure highlights observations from 17:39 UT to 18:20 UT, capturing mostly the impulsive phase of the flare, where IRIS and AIA see the development and evolution of the flare ribbons. 

We use the SDO AIA $1600$ \AA{} intensity maps and HMI magnetic field measurements to derive reconnection fluxes and reconnection rates {as described in Equation~\ref{eq:flux}}. {The $1600$ \AA{} intensity maps correspond to a characteristic temperature response of  $5000$ K and $10^5$ K, sensitive to chromospheric emission (mainly contain C IV doublet 1548.20 \AA{} \& 1550.77 \AA{}, and continuum emission). AIA $1600$ \AA{} data has a spatial resolution of $0.6$\arcsec\ and  cadence of $24$ seconds. HMI vector magnetic field maps, which include the line-of-sight (LOS) field strength, inclination and azimuthal angles,  are transformed into the vector magnetic field components in the local reference frame B$_{x}$, B$_{y}$, and B$_{z}$ \citep{Sun2013OnNote,Bobra2021Mbobra/SHARPs:2021-07-23}.} The HMI magnetic field data {has} a $720$ second cadence and $0.5$\arcsec\ pixel resolution \citep{Kazachenko2017}. In the local reference frame, B$_{z}$ is the magnetic field component perpendicular to the active region solar surface. AIA observed the entire evolution of the flare ribbons within a full-disk field-of-view (FOV). {The} IRIS Slit-Jaw Imager (SJI) provides higher resolution images with $0.33$\arcsec\ spatial pixel resolution (almost twice that of AIA 1600 \AA{}) and a cadence of 17 seconds in the 1330 \AA{} passband within a limited FOV (the dashed green box in the AIA and HMI maps of Figure~\ref{fig:flare_cont} shows the IRIS-SJI FOV). The 1330 \AA{} SJI channel probes from  {coronal} to chromospheric temperatures ($10^7 -5000$ K; {which mainly contains C II doublet 1334 \AA{} \& 1335 \AA{}, Fe XII 1349.4 \AA{}, Fe XXI 1354.08 \AA{},} and continuum emission) and reveals additional spatial structures of the flare ribbons. 

To describe the non-thermal and high-energy flare properties, we use observations from the IRIS Spectrograph (SG), Fermi’s \textit{Gamma-ray Burst Monitor} (GBM, \citealt{Meegan2009TheMonitor}), and GOES X-ray Sensor (XRS; \citealt{Machol2019GOES-RIrradiance}). IRIS-SG measures spectral profiles from multiple emission lines, including the Si IV $1402.77$ \AA{} transition region line. We fit a single Gaussian profile to the Si IV spectra to calculate intensity, Doppler velocity, and non-thermal velocity ($\nu_\text{non-thermal}$) within the raster region (blue rectangle in Figure~\ref{fig:flare_cont}) every 34 seconds. We use GBM's Continuous Spectroscopy (CSPEC) data, downloaded with \textit{OSPEX} IDL software\footnote{\url{https://hesperia.gsfc.nasa.gov/fermi\_solar/analyzing\_fermi\_gbm.htm}}, to evaluate HXR temporal and spectral evolution during flares. The CSPEC data has a varying cadence from $1.0$ seconds to $32.7$ seconds, and a default value of 4.0 seconds, and is  stored in 128 quasi-logarithmic energy bins, which can be integrated into the following energy channels of interest: $8-15$ keV, $15-25$ keV, $25-50$ keV, $50-100$ keV, and $100-300$ keV. Finally, to measure flare lightcurves in soft X-ray (SXR), we use observations from GOES XRS with a cadence of $1$ s in two channels, $0.5 - 4$ \AA{} and $1 - 8$ \AA{}, corresponding to an energy range between about $1.5$ and $25$ keV.    

\subsection{Tracking the Flare Ribbon Bright Leading Edge (FRBLE)}\label{subsec:met_FRBLE}
To describe the small-scale  morphological changes of the ribbon in the plane-of-sky (POS), we decompose the intensity images from IRIS ($I$) into various features. Appendix~\ref{app:FRBLE} shows the intensity decomposition and feature extraction application to a synthetic image. This application to synthetic data illustrates the general usability of our feature extraction methodology to astrophysical image datasets. First, we distinguish flare-ribbon emission ($I_{\text{FR}}^*$) from background emitting plasma ($I_{\text{BG}}^*$). We use intensity decomposition on a logarithmic scale ($I^* =  \log_{10}(I)$) to allow for a more easy visualization of high-contrast spatial features. Extracting the flare ribbon from the background emission can be achieved with many different methods (e.g., \citealt{Yang2018AutomatedImages,Kirk2013AnBrightenings,Maurya2010ARelease}). We use the \cite{Qu2003AutomaticSVM,Qiu2004MagneticEvents} intensity threshold method with
\begin{align}
    I^*(\imgvar) =  & 
    \begin{cases}
        I^*_{\text{FR}}(\imgvar;\epsilon_{\text{FR}})  \text{ if } {I}^*(\imgvar) \in {I}^*(\imgvar)\mathcal{N}(\imgvar;c;\tau_{\text{FR}};\epsilon_{\text{FR}}),\\ 
         I^*_{\text{BG}}(\imgvar;\epsilon_{\text{FR}}) \text{ if } {I}^*(\imgvar) \notin {I}^*(\imgvar)\mathcal{N}(\imgvar;c;\tau_{\text{FR}};\epsilon_{\text{FR}}).
    \end{cases}    
\end{align}
Here we define an instantaneous threshold mask $\mathcal{N}$ for position ($x_\text{i}$, $y_\text{j}$) at observation time $t_\text{k}$ as 
\begin{align}
    \mathcal{N}(\imgvar;c;\tau_\text{FR}; \epsilon_{\text{FR}}) = & 
    \begin{cases}
        1 \text{ if } \bar{I}^*(\imgvar;\tau_\text{FR}) + \epsilon_{\text{FR}}(\imgvar;\tau_\text{FR}) \geq c*\tilde{I^*},\\ 
        0 \text{ if } \bar{I}^*(\imgvar;\tau_\text{FR}) + \epsilon_{\text{FR}}(\imgvar;\tau_\text{FR})< c*\tilde{I^*}.\\
    \end{cases}
\end{align}
In these equations $\mathcal{N}$ discriminates among pixels above the constant $c$ multiplied by the median intensity $\tilde{I^*}$ for all observation frames. The value of $c$ is empirically chosen so that $\mathcal{N}$ best matches the spatial distribution of $I^*_{\text{FR}}$. Typical $c$ values used for the 1330 \AA{} and 1400 \AA{} SJI channels are 8 on the linear scale and 1.7 on the logarithmic scale. {The intensity at each point in the frame is substituted by the running-window temporally averaged intensity $\bar{I}^*$, where $\tau_\text{FR} = \pm95$ s (10 consecutive IRIS-SJI frames) is the chosen temporal window. We choose $\tau_\text{FR} = \pm95$ s empirically because it eliminates bright transients produced by heating episodes spatially uncorrelated to the flare ribbon emission while maintaining the original spatial distribution of the flare ribbons. Temporally averaging the intensity reduces the occurrence frequency of transient bright features contaminating the instantaneous flare ribbon mask.} Errors in the masking procedure due to thresholding, transient bright structures, pixel saturation and bleeding, dust on top of the camera detector, pixel interpolation, and other sources are represented by the variable $\epsilon_{\text{FR}}$. \cite{Yang2018AutomatedImages} have indicated that this instantaneous flare ribbon masking method ($\mathcal{N}$) can overestimate the area enclosing the flare ribbon and, therefore, would fail to accurately track the morphological changes of the flare ribbon substructure. {Furthermore, as discussed by \cite{CorchadoAlbelo2024InferringRates}, the cumulative mask {($\mathcal{M}$)} {indicates} the first time a given pixel intensity $I(\imgvar$) exceeds the chosen background threshold and thus {corresponds to} the first brightening of the flare ribbon at a given coordinate ($\mathcal{M}(\imgvar) = \mathcal{N}(\imgvar) \cup \mathcal{M}(\imgvartb)$)}. Moreover, substructures have been observed to develop in regions within the cumulative mask of the flare ribbon, making this method ill-suited for {temporally} tracking the substructure development on flare ribbon observations. {Therefore, we modify the methodology of \cite{Qiu2004MagneticEvents} to track the evolution of substructures, and limit the use of the cumulative ribbon mask method to calculate the reconnection flux and rate.} 

The substructure related to current sheet processes develops in the apparent interface between the leading edge and the rest of the flare ribbon body \citep{Kerr2021HeIonizations,Naus2022CorrelatedFlare,Polito2023SolarSpectroscopy,Kerr2024SolarFootpoints,Pietrow2024SpectralRibbons}. We call this interface, which corresponds to the brightest region of the flare ribbons, the Flare Ribbon Bright Leading Edge (FRBLE). {In the context of ultraviolet flare kernels observed in the IRIS-SJI channels \citep{Kowalski2017The29}, the FRBLE is the area enclosing all of the kernels. Furthermore, we use the cumulative area enclosed by the FRBLE to calculate the reconnected magnetic flux in observations 
\begin{equation}
    \Phi^{\pm} = \int\int_{\mathcal{M}(\mathcal{B_{\text{FRBLE}}})} \text{B}^{\pm}_z dA_\text{FRBLE}, 
\end{equation}
from which we then take the derivative to find the reconnection flux rate    \citep{Kazachenko2017,Naus2022CorrelatedFlare,CorchadoAlbelo2024InferringRates}.} Trailing behind this region is the so-called flare ribbon moss, which forms a web-like structure of all cooling flare ribbon pixels \citep{Fletcher2001TheRibbons}. This moss region also has substructure, but is believed to arise {from the} energy dissipation in the lower atmospheric layers and therefore does not serve as a diagnostic of the flare current sheet conditions. To identify the spatial substructure, we decompose flare ribbons into moss and FRBLE areas:
\begin{align}
    I^*_\text{FR}(\imgvar;c) = & 
    \begin{cases}
        I^*_{\text{FRBLE}}(\imgvar;c;l)  \text{ if } I^*_\text{FR}(\imgvar;c) \in I^*_\text{FR}(\imgvar;c)\mathcal{L}(\imgvar;c;l),\\ 
        I^*_{\text{Moss}}(\imgvar;c;l) \text{ if } I^*_\text{FR}(\imgvar;c) \notin I^*_\text{FR}(\imgvar;c)\mathcal{L}(\imgvar;c;l),
    \end{cases}    
\end{align}
where $\mathcal{L}$ is a FRBLE mask
\begin{align}
   \mathcal{L}(\imgvar;c;l) = &
   \begin{cases}
        1 \text{ if } \mathcal{N}(\imgvar;c)\nabla^2( I^*(\imgvar)) \leq l, \\ 
        0 \text{ if } \mathcal{N}(\imgvar;c)\nabla^2( I^*(\imgvar)) > l.\\
    \end{cases}
\end{align}
Here we define the FRBLE region as the area with a negative second spatial derivative ($\nabla^2I^*$), in the logarithmic scaled image, below a given threshold $l = -10^{-2}$ pixel$^{-2}$. We use the Laplacian of Gaussian (LoG) filter from the Python Scipy image processing package (\citealt{Virtanen2020SciPyPython,Marr1980TheoryDetection}) to estimate $\nabla^2I^*$. The LoG operator has been used by \cite{Gill2010UsingRibbons} to detect the flare ribbon leading edges on AIA 1600 \AA{} observations. We use a standard deviation of $\sigma = 2.5$ pixels $\approx 0.603$ Mm to define the Gaussian function used to reduce the noise in the $\nabla^2I^*$ calculation while {avoiding} degrading substructures.

{Finally, since the extracted mask can include small artifacts, we clean the extracted mask.} We eliminate {the} small bright artifacts, arising from numerical noise and transient brightening in the 1300 \AA{} channel, by imposing that the length ($\mathcal{P}$) of each $\mathcal{L}$ segment be greater than $10\%$ of the largest segment. {Afterwards, we use the clean FRBLE binary image to extract the boundary $\mathcal{B}_{\text{FRBLE}}(\imgvar;c;l)$, using the OpenCV Python package (\citealt{Bradski2000TheLibrary}). We then used $\mathcal{B}_{\text{FRBLE}}$ to infer the spatial complexity along the FRBLE and track the substructures of interest.}

\subsection{Quantifying the Spatial Complexity of Flare Ribbon Substructure}\label{subsec:met_CDM}
Flare ribbon substructures that have been linked to coronal plasmoids (e.g., swirls and wave-like features) can be mathematically described as small-scale morphological variations of the FRBLE boundary,  $\mathcal{B}_{\text{FRBLE}}$. Therefore, to track the development and evolution of {the} flare ribbon substructure, we seek a metric that quantifies the morphological complexity of the boundary across multiple spatial scales. Appendix~\ref{app:CDM} and Figure 2 of \cite{Mason2022StatisticalBoundary} show application of such morphology complexity metric to curvilinear and multi-scale boundaries. {These tests and example figures illustrate the general applicability of the metrics and its sensitivity to the selected range of spatial scales.} 

Flare ribbons often exhibit complex large-scale curvilinear morphology (e.g. \textit{J-shaped} flare ribbon hooks; \citealt{Janvier2016EvolutionFlare}), caused by the formation of newly reconnected loops and the global magnetic topology described by the three-dimensional standard flare model \citep{Longcope2007Modeling2004, Aulanier2013TheDimensions, Janvier2013TheDimensions}. {To isolate the complexity changes associated with flare-ribbon substructure evolution and potential diagnostic of current-sheet fragmentation, we restrict our analysis to spatial scales smaller than those characterizing the large-scale ribbon morphology.} Furthermore, since flare ribbons in observations are not inherently contiguous structures, and numerical noise may introduce artificial fragmentation in $\mathcal{L}$, each ribbon segment is analyzed separately to measure local spatial complexity along $\mathcal{B}_{\text{FRBLE}}$. For this purpose, we use the spatially dependent correlation integral ($C(r,x,y)$; \citealt{Grassberger1983MeasuringAttractors}) to perform Correlation Dimension Mapping (CDM), following \cite{Mason2022StatisticalBoundary}:  
\begin{align}
    C(r,x,y,t) = & \sum_{\text{i}=1}^{N(x,y,t)} \Theta\left(r -\sqrt{\left( x_\text{i}-x\right)^2 + \left( y_\text{i}-y\right)^2}\right),\\ 
    C(r,x,y,t) \sim & r^{\mathcal{D}(x,y,t)}.
\end{align}
Here $C$ is the spatially dependent correlation integral \citep{Grassberger1983MeasuringAttractors}, $\Theta$ is the Heaviside function (\citealt{Heaviside2011ElectricalPapers}, defined as 1 when its argument is positive and 0 otherwise), $r$ is the spatial scale of interest, and $N$ is the number of points along the boundary within the maximum radial scale $r_{\text{max}}=\text{max}(r)$, at time $t$. The spatially dependent correlation dimension $\mathcal{D}(x,y,t)$ is obtained by fitting a linear relation to the log–log dependence of $C(r,x,y,t)$ on $r$ over the range $[r_{\text{min}}, r_{\text{max}}]$. 

To characterize the overall complexity of the FRBLE boundary, we define a weighted mean correlation dimension, 
\begin{align}
\bar{\mathcal{D}} = \sum^{N_\text{FRBLE}}_i \mathcal{D}_iw_\mathcal{D},
\end{align}
where $w_\mathcal{D} = N_{\mathcal{D}}/N_\text{FRBLE}$ represents the fraction of boundary points with correlation dimension $\mathcal{D}$, and  $N_{\mathcal{D}}$ is the total number of points along the FRBLE boundary.

The correlation dimension is conceptually related to {the Minkowski–Bouligand dimension (\citealt{Minkowski1901UeberVolumen., Bouligand1929SurPlan})}, more commonly referred to as the {\bf box-counting dimension} ($\mathcal{D}_\text{BC}(N,\varepsilon,t)$), of a binary image. The box-counting dimension is defined as
\begin{align}
\mathcal{D}_\text{BC}(N,\varepsilon,t) = -\lim_{\varepsilon \to 0} \frac{\log N(\varepsilon,t)}{\log\varepsilon},
\end{align}
where $N(\varepsilon,t)$ denotes the number of boxes with linear size $\varepsilon$ needed to cover the area of a binary object at time $t$. A fractional value of $\mathcal{D}_\text{BC}$ describes the global fractal, or space filling properties of the object's boundary. 

Importantly, the box-counting dimension $\mathcal{D}_\text{BC}$ provides a global measure of spatial complexity and  does not resolve localized variations along the FRBLE boundary $\mathcal{B}_{\text{FRBLE}}$ that are used to track substructure evolution. {In our application, $\mathcal{D}_\text{BC}$ describes the overall FRBLE's global spatial-complexity at each magnetic polarity, incorporating the full spatial extent of the boundary morphology.} 

In contrast to the box-counting method, the CDM method provides a spatially-localized complexity value at each boundary point $\bar{\mathcal{D}}$, computed over a neighborhood determined by the selected scale range. This enables the identification of specific regions along the FRBLE potentially impacted by the flare reconnection dynamics. Furthermore, unlike standard box-counting, CDM can be applied to any type of boundary that is non-fractal, fractal, or multi-fractal in nature. 

\section{Results}\label{sec:res}
In this section, we apply the Flare Ribbon Bright Leading Edge (FRBLE) tracking algorithm to IRIS 1330 \AA{} SJI data of the 2015 June 22 M6.5 flare and quantify the spatial complexity of the evolving ribbon structure using the Correlation Dimension Mapping (CDM) and Box-Counting Dimension. 

\subsection{Spatial Evolution of Flare Ribbon Substructure Complexity}\label{subsec:spatFRBLE}
\begin{figure}
    \centering
    \includegraphics[width=\linewidth,trim={6cm 3cm 9cm 2cm}, clip]{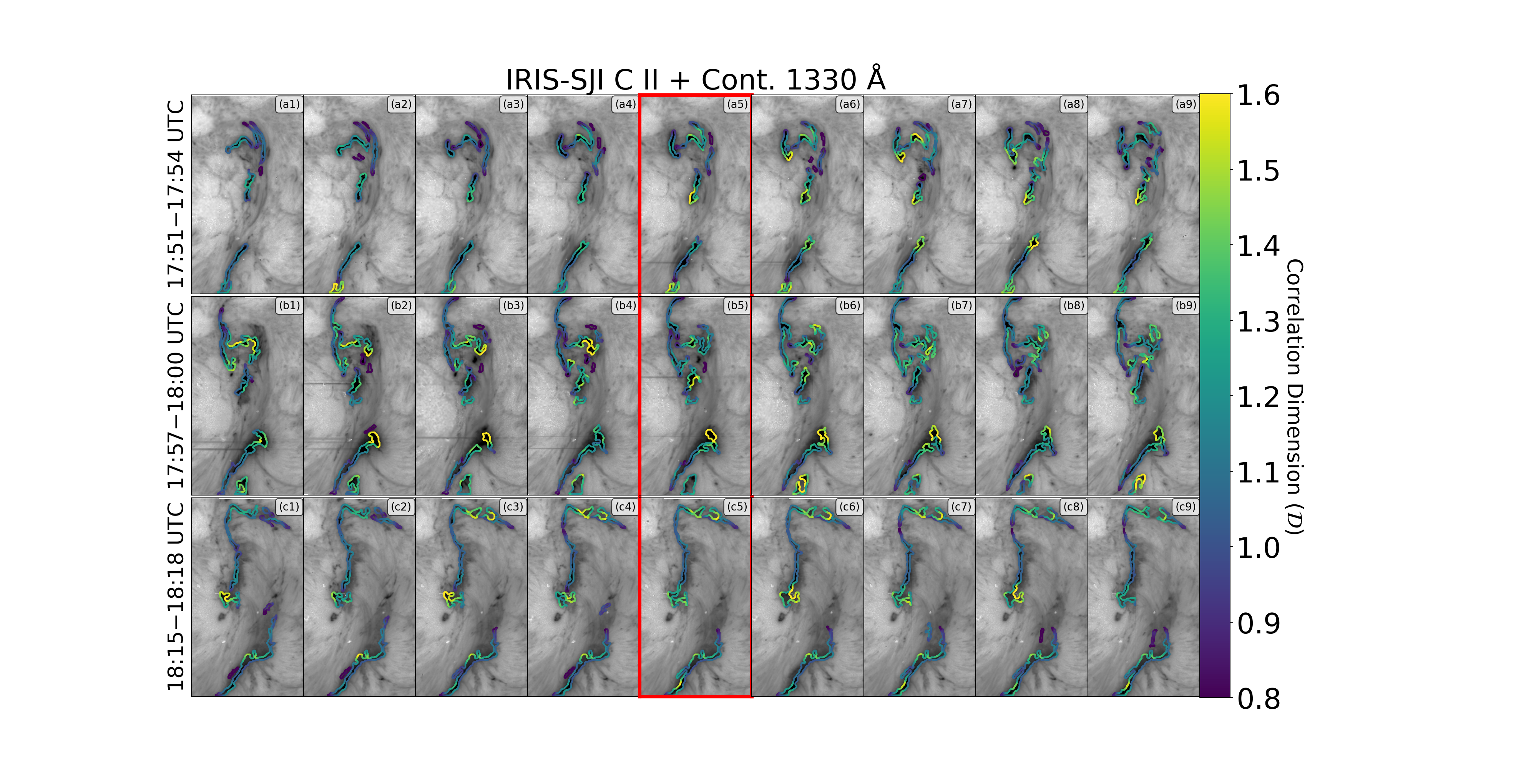}
    \caption{{Spatial evolution of the FRBLE boundaries and local correlation dimension ($\mathcal{D}$) over IRIS 1330~\AA\ SJI. Images are shown as $\log_{10}$ intensity ($I^*$; 1--$10^{4}$ DN; reversed grayscale). Rows (a1-a9), (b1-b9) and (c1-c9) show nine consecutive IRIS frames, during 17:51--17:53 UT, 17:57--18:00 UT and 18:15--18:18 UT, respectively. Red boxes mark the middle frames: 17:52 UT (a5), 17:58 UT (b5), and 18:16 UT (c5). Colors show $\mathcal{D}$ along the FRBLE: higher values (yellow; $\mathcal{D}>1.5$) indicate more corrugated/complex segments, while lower values (blue/dark; $\mathcal{D}\le 1$) indicate smoother segments.}}
    \label{fig:res_Spat_CDM}
\end{figure}

Figure \ref{fig:res_Spat_CDM} shows the extracted FRBLE at different stages of the flare impulsive phase (Figure~\ref{fig:flare_cont}(a) SXR light curve). The middle red column highlights the different stages of the 1330 \AA{} emission: (a5) onset of the rapid increase ($t \approx$ 17:52 UT), (b5) the peak ($t \approx$ 17:58 UT), and the decay ($t \approx$ 18:16 UT). The IRIS slit-jaw imager (SJI) FOV captures the full evolution of the northernmost ribbon, mostly embedded in the positive vertical magnetic field polarity (see Figure~\ref{fig:flare_cont}). However, the southernmost ribbon, associated with the negative polarity, is only partially captured within the IRIS-SJI FOV, representing a limitation of our analysis. The missing segment of the southernmost ribbon corresponds primarily to the \textit{J-shaped} hook structure of the negative magnetic polarity. Such hook structures are generally interpreted as the chromospheric footpoints of the erupting flux rope (e.g. \citealt{Demoulin1996ThreedimensionalTubes, Janvier2013TheDimensions, Janvier2016EvolutionFlare,Savcheva2012SIGMOIDALMODEL,Savcheva2015THERIBBONS,Savcheva2016THEEVOLUTION}). This interpretation is consistent with observations of a halo CME \citep{Gopalswamy2018Sun-to-earthObservations} associated with this M6.5 flare. 

As seen in Figure~\ref{fig:res_Spat_CDM}, fine substructure develops through the {entire} extent of the observed flare ribbon. This behavior is qualitatively consistent with theoretical interpretations in which the FRBLE substructure reflects the properties of the three-dimensional reconnection current sheet \citep{Wyper2021IsSheet, Dahlin2025DeterminingReconnection}. The incomplete coverage of the negative-polarity hook therefore suggests {that our analysis may underestimate the substructures associated with magnetic field lines rooted in the erupting flux rope's topological domain. Nevertheless, high-spatial resolution observations of two-flare-ribbons rarely capture the full spatial extent of the ribbon kernels in both B$_z$ polarities. Even with partial coverage, our case study therefore represents a realistic application of the FRBLE tracking and CDM analysis, which can be systematically repeated for other high-resolution flare ribbon observations.}

Throughout the impulsive phase of the flare, the FRBLE boundary exhibits a modest degree of fragmentation. This effect is most significant during the early and late stages of the impulsive phase (top and bottom rows of Figure~\ref{fig:res_Spat_CDM}). {Early in the impulsive phase ($\text{17:38 UT}\lesssim \text{t} \lesssim \text{17:51 UT}$), the flare ribbons develop from a few compact bright kernels (somewhat indistinguishable from other transient bright features) into a more contiguous distribution of ribbon kernels. Toward the flare decay phase ($\text{t} \gtrsim \text{18:14 UT}$), the HXR emission in the $25 - 50$ keV (see Figure~\ref{fig:flare_cont}) and microwave emission ranging from $4.43-16.9$ GHz (see Figure 1 in \citealt{Kuroda2018Three-dimensionalFlare}) both reach a stable minimum. At this later stage, the flare-ribbon kernels progressively fade, increasing the difficulty of separating FRBLE emission from background moss structures in the observations.} The apparent fragmentation of the FRBLE boundary at these times is beyond the scope of this study. Instead, we focus on describing the fine-scale substructure along the leading edge segments, which are independently analyzed using the CDM (see Section~\ref{subsec:met_CDM}). 

For the spatial scales of the CDM procedure we use a range from a minimum radius of $3.327$\arcsec\ (10 pixels $\approx 2.413$ Mm) to a maximum radius of $4.991$\arcsec\ (15 pixels $\approx 3.619$ Mm). This range is constrained by {the} LoG operator used in the FRBLE extraction procedure. In particular, the lower limit is set by the Nyquist-Shannon sampling theorem ($r_\text{min} = 4\sigma = 10 \text{ pixels}$; \citealt{Nyquist1928CertainTheory,Shannon1949CommunicationNoise}). The upper limit of spatial scales is manually selected to capture the smallest corrugations along the boundary while {minimizing} the inclusion of larger curvilinear structures in the main ribbons. As shown in Figure~\ref{fig:res_Spat_CDM}, most of the FRBLE boundary exhibits a relatively low correlation dimension, with $1.0 \lesssim \mathcal{D} \lesssim 1.2$. Regions with the highest correlation dimension $\mathcal{D}$ highlight areas where the FRBLE boundary {is highly corrugated; such regions appear intermittently along the entire boundary}. These localized enhancements in spatial complexity are short-lived, persisting for approximately $19~\mathrm{s} \lesssim \Delta t \lesssim 95~\mathrm{s}$  before decaying as the FRBLE becomes less spatially complex. 

\subsection{Temporal Evolution of Flare Ribbon Substructure Complexity} \label{subsec:timeFRBLE}

\begin{figure}
    \centering  \includegraphics[height=0.5\textheight,trim={0cm 0cm 0cm 0cm}, clip]{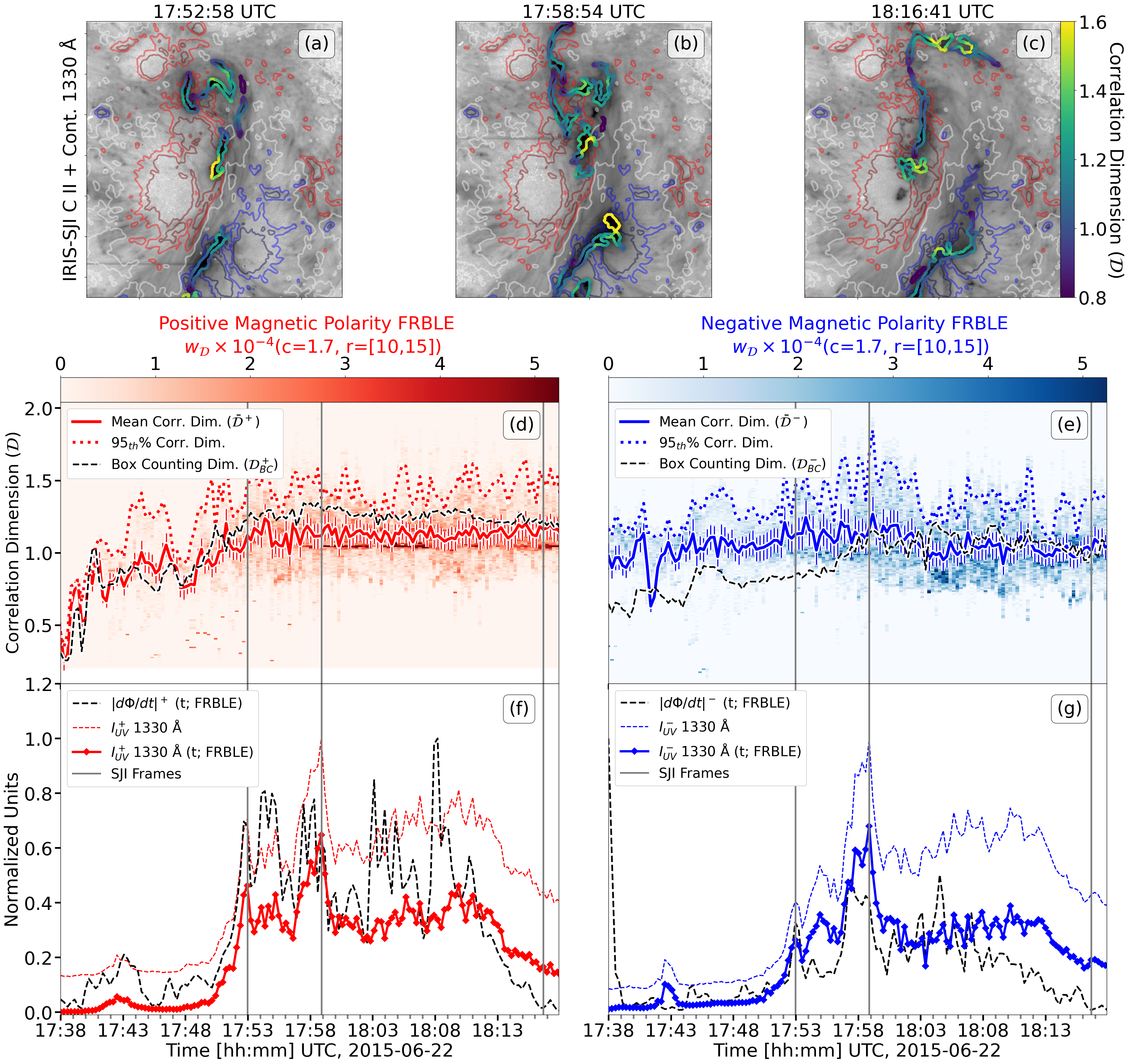}
    \caption{{Evolution of FRBLE complexity and reconnection proxies. (a--c) Example frames (same as Fig.~\ref{fig:res_Spat_CDM} a5, b5, c5) showing $\mathcal{D}$ along the FRBLE at 17:52, 17:58, and 18:16~UT. Contours show HMI $B_z$ levels: +500/+1000~G (red), $-500/-1000$~G (blue), and 0~G (white). (d,e) Distributions $w_{\mathcal{D}}$ for the positive and negative polarity ribbons. Solid lines show the mean $\bar{\mathcal{D}}$, dotted lines the 95th percentile, and black dashed lines the box-counting dimension $\mathcal{D}_{\rm BC}$. (f,g) Time series reconnection dynamics proxies in positive and negative magnetic-polarity FRBLEs, respectively. The solid and dashed colored-lines (red and blue) represent the 1330 \AA{} intensity emitted from the FRBLE and full SJI FOV respectively, while the black-dashed line shows the cumulative reconnection rate derived from FRBLE observations. Vertical gray lines mark the times in (a--c).}}
    \label{fig:res_CDM_hist}
\end{figure}

Although the spatial properties of individual high-complexity regions are of interest, in this study we focus primarily on the global evolution of FRBLE complexity. {In Figure~\ref{fig:res_CDM_hist}(a--c), we overview the spatial distribution of values of FRBLE complexity, described by $\mathcal{D}$, along the FRBLE boundary}. The northern (top) ribbon spans the positive polarity footpoints of the newly formed flare arcade, whilst the southern (bottom) ribbon spans the negative polarity. 

In {Figures~\ref{fig:res_CDM_hist}(d,e) we show the evolution of the statistical weight spatial complexity metric $w_\mathcal{D}$ separately within each magnetic field polarity.} We analyze each polarity separately to determine whether observed trends in ribbon morphology correspond to similar reconnection processes in the current sheet. During the early impulsive phase at $17{:}38~\mathrm{UT}-17{:}48~\mathrm{UT}$, the positive-polarity FRBLE consists of a small number of compact kernels whose spatial extent is comparable to the CDM scale range. {Because of the comparable CDM scale range and lack of well-developed substructure, $w_\mathcal{D}$ at these times concentrates in the positive polarity at lower correlation dimensions ($\mathcal{D}< 1$, see Appendix~\ref{app:CDM} for details). In contrast, the negative polarity FRBLE is larger compared to the CDM spatial scales, and exhibits more complex features from onset, yielding correlation dimensions closer to unity ($\mathcal{D} \approx 1$)}. {The overall evolution of $w_\mathcal{D}$ in both polarity ribbons is well represented by the mean correlation dimension ($\bar{\mathcal{D}}$), which we adopt as the primary metric describing global complexity evolution. To track the most spatially complex regions, we use the $95^{\text{th}}$ percentile of $w_{\mathcal{D}}$ (dotted lines in Figures~\ref{fig:res_CDM_hist}(d,e)). High-complexity regions ($\mathcal{D}> 1.1$) in the positive polarity at this early stage temporally coincide with a small burst in the reconnection rate and intensity enclosed within the FRBLE.} 

Beginning around 17:53~UT, both polarities show a shift of $w_{\mathcal{D}}$ toward higher spatial complexity ($\mathcal{D} > 1$).  This interval corresponds to bursty emission observed in HXR, SXR, microwave wavelengths, together with peaks in the reconnection rates derived within FRBLE areas. These signatures {were} previously reported as multi-wavelength quasi-periodic pulsations rooted in the FRBLE region (see Figure~2 \& 6 of \citealt{Kuroda2018Three-dimensionalFlare}). At times of local intensity maxima in the 1330~\AA\ emission and reconnection rate (black dashed vertical lines), small but systematic increases in $\bar{\mathcal{D}}$ are observed, driven by enhanced high-percentile complexity. 

After around 18:00 UT, both the reconnection rate and $1330$ \AA{} emission decrease significantly, accompanied by a modest decline in FRBLE complexity. This level of complexity is maintained even as the flare approaches the decay phase, and the {length of the FRBLE begins to decrease} (Figure~\ref{fig:res_CDM_hist}(c)). To quantify this relationship, we compute Spearman rank correlation coefficients between the reconnection flux rate and the mean correlation dimension in positive and negative $B_z$ polarities (\citealt{Spearman1904TheThings}). We find that the correlation coefficient is moderate: $r_S(|d\Phi/dt|^{\pm}, \bar{\mathcal{D}}^{\pm}) = \{0.4, 0.3\},$
in positive and negative $B_z$ polarities, respectively (\citealt{Spearman1904TheThings}). 

\begin{figure}
    \centering
    \includegraphics[width=\linewidth,trim={1.6cm 1.3cm 3cm 1cm}, clip]{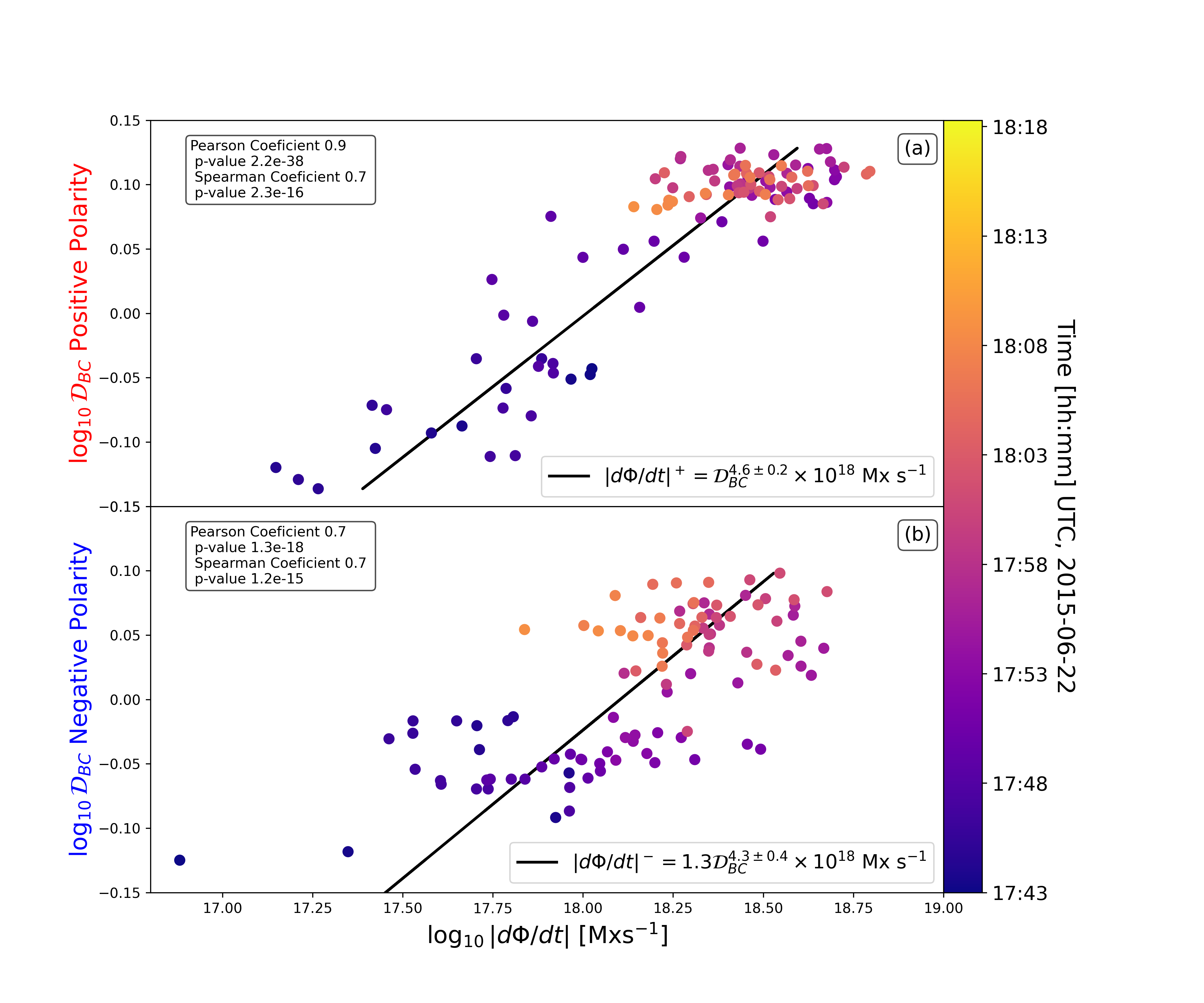}
    \caption{{Reconnection rate vs. box-counting dimension. Absolute reconnection flux rate $|d\Phi/dt|$ versus box-counting dimension $\mathcal{D}_{\rm BC}$ for the (a) positive and (b) negative polarity FRBLEs. Dots' color encodes time (from $\sim$17:43~UT to $\sim$18:13~UT). Pearson and Spearman coefficients are listed in each panel. Best-fit power-law scales are shown in the lower right of each panel.}}
    \label{fig:log-log_corr}
\end{figure}

In Figure~\ref{fig:log-log_corr} we compare the box-counting dimension with FRBLE-derived reconnection flux rate in both polarities, from $\sim$17:43 UT to $\sim$18:13 UT, finding a power law relationship between the two: $|d\Phi/dt|^{+} \propto \mathcal{D}_\text{BC}^{4.6\pm0.2}$ and $|d\Phi/dt|^{-} \propto \mathcal{D}_\text{BC}^{4.3\pm0.4}$. Furthermore, we find a strong correlation between the temporal evolution of box-counting dimension in the two polarities  ($r_S(|d\Phi/dt|^{\pm},\mathcal{D}_\text{BC}^\pm)=0.7$).

\begin{figure}
    \centering  
    \includegraphics[width=\linewidth, trim={5cm 1cm 5cm 2cm}]{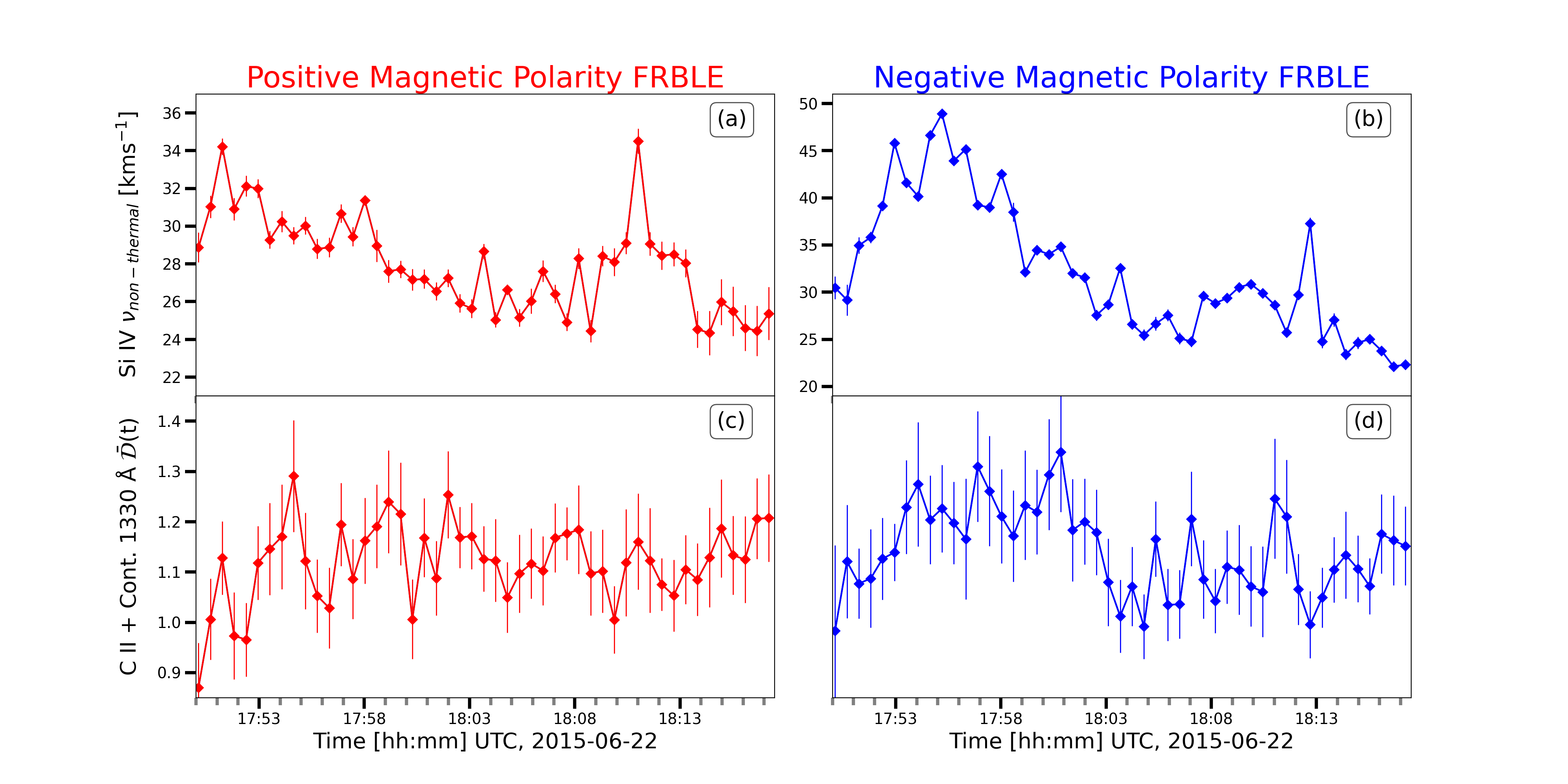}
    \caption{{Mean non-thermal velocity vs. mean FRBLE complexity within the IRIS-SG raster region. (a,b) Mean Si IV non-thermal velocity $\bar{\nu}_{\rm non\text{-}thermal}$ in the positive (red) and negative (blue) polarity FRBLE regions inside the raster region (blue rectangle in Fig.~\ref{fig:flare_cont}).
    (c,d) Corresponding mean correlation dimension $\bar{\mathcal{D}}$ for the same regions and color scheme.
    }}
    \label{fig:res_CDM_Rast}
\end{figure}

As an independent test of the relationship between ribbon substructure and reconnection dynamics, in Figure~\ref{fig:res_CDM_Rast}, we compare the evolution of the mean non-thermal velocity $\bar{\nu}_\text{non-thermal}(t)$ with the mean correlation dimension of the FRBLE $\bar{\mathcal{D}}(t)$. Each data point is recorded at the mean time of the 16 slit scans (every 34 seconds) composing each raster image. Importantly, the start time of Figure~\ref{fig:res_CDM_Rast} is a subsample of the entire time series, starting when the FRBLE region first appears in the slit-raster region at t $\approx$ 17:50 UT. At early times, the non-thermal velocity peaks in both the positive ($35$ km s$^{-1}$ at t $\approx$ 17:52 UT) and negative ($48$ km s$^{-1}$ at t $\approx$ 17:55 UT) polarity ribbons, coinciding with the increase in 1330 \AA{} emission and the onset of bursty reconnection rates in both ribbons (Figure~\ref{fig:res_CDM_hist}). After t $\approx$ 18:01 UT, both ribbons show a decrease in non-thermal velocities, followed by a final burst around 18:12 UT. We find a moderate correlation between the mean correlation dimension and non-thermal velocity within the negative FRBLE within the IRIS-SG observed region ($r_S(\bar{\mathcal{D}}^-,\nu^{-}_\text{non-thermal})=0.4$). In contrast, in the positive polarity IRIS-SG region, we find no significant monotonic correlation ($r_S(\bar{\mathcal{D}}^+,\nu^{+}_\text{non-thermal})=-0.2$).

\section{Discussion}\label{sec:dis}
In this study, we track the multi-spatial scale evolution of the flare ribbon leading edge {boundary} (FRBLE) and evaluate its relationship to observational proxies of magnetic reconnection processes. Our working hypothesis is that FRBLE boundary complexity reflects the underlying reconnection dynamics (Appendix~\ref{app:MultiScaleArg}; \citealt{Forbes1984NumericalRegion,Forbes2000WhatEjections}). 

Using the box-counting dimension, we first quantify the {\it global} multi-scale morphology of the FRBLE. We find a strong monotonic correlation between the box-counting dimension and the reconnection flux rate in each magnetic polarity 
\begin{align}
r_S(|d\Phi/dt|^{\pm},{\mathcal{D}_\text{BC}^{\pm}})=\{0.7,0.7\}
\end{align}
during 17:43 UT $\lesssim$ t $\lesssim$ 18:13 UT). {During} the same time range{,} the two variables exhibit a power law relationship 
\begin{align}
d\Phi/dt \propto{\mathcal{D}_\text{BC}^{\gamma}},
\end{align}
(see Figure~\ref{fig:log-log_corr}). {From cross-correlation analysis we find that reconnection flux rate lags the box-counting dimension by one time step (approximately 17~s, averaged across polarities).} Given the cadence and uncertainties associated with computing the temporal derivative of magnetic flux, this delay is consistent with near co-temporal behavior, suggesting that the multi-spatial-scale ribbon morphology closely tracks reconnection dynamics.

In Appendix~\ref{app:MultiScaleArg} we interpret the relationship between box-counting dimension and the magnetic reconnection flux rate (see Figure~\ref{fig:log-log_corr}) approximated from FRBLE evolution as evidence of a fragmented current sheet. One mechanism that links the {multi-spatial-scale nature of the FRBLE boundary to a fragmented coronal current sheet is} the 3D oblique plasmoid instability (\citealt{Wyper2021IsSheet, Dahlin2025DeterminingReconnection}). With a parameterized analytical model \cite{Wyper2021IsSheet} recently found that current sheet plasmoids (sunward and anti-sunward) trace to sub-scale deformations (wrappings, swirls, wave-like structures) along the flare ribbons. This result has been validated by \cite{Dahlin2025DeterminingReconnection} using the Adaptively Refined Magnetohydrodynamics Solver (ARMS; \citealt{DeVore2008HomologousBreakout}). However, these models do not include physical treatment of the particle energy gained due to non-thermal processes (\citealt{Drake2006ElectronReconnection,Drake2010ARAYS,Guidoni2016MAGNETIC-ISLANDFLARES,Dahlin2017TheReconnection,Huang2017DevelopmentIsland,Guidoni2022SpectralIslands}), nor the energy transport that gives rise to flare ribbon emission \citep{Ramaty2000HighFlares,Fletcher2008,Brosius2012EXTREME-ULTRAVIOLETFLARE,Reep2016ALFVENICFLARES,Polito2023SolarSpectroscopy}. Regardless, the dynamics (and presence) of plasmoids in these {3D} simulation results in a multi-spatial-scale flare leading edge boundary. Another fragmented current sheet scenario recently proposed is turbulent 3D reconnection \citep{Wang2023Three-dimensionalSheet,Wang2025BasicSheets}. In these 3D simulations the turbulent current sheet breaks up into multi-spatial scale reconnection kernels, with spatial-scale-invariant reconnection rates. While our observational analysis suggests that magnetic reconnection occurs in a fragmented current sheet during this M-class flare (2015-06-22), it cannot distinguish the underlying physical details of the reconnection (e.g. plasmoid mediated reconnection, turbulent reconnection). 

We then examine {\it localized} ribbon substructure using the correlation dimension metric (CDM, \citealt{Mason2022StatisticalBoundary}). Our CDM analysis shows that high-complexity regions are spatially localized and short-lived, persisting for approximately $19 \text{ s}\lesssim \Delta t \lesssim 95\text{ s}$ after which they disperse or smooth at the analyzed substructure scales (2.4 - 3.6 Mm). We find that the mean CDM complexity metric along each magnetic polarity FRBLE exhibits a moderate correlation with the corresponding reconnection rate ($r_S(|d\Phi/dt|^{\pm},\bar{\mathcal{D}}^{\pm}) = \{0.4,0.3\}$ (see Figure~\ref{fig:res_CDM_hist}) {and precedes it  by approximately seven time steps ($\sim$119~s).} This temporal offset and moderate correlation suggests that localized substructure evolution may not directly trace instantaneous reconnection rate variations.

To further explore this possibility, we compare the LOS non-thermal velocity with the evolution of the {CDM complexity metric measured on the POS.} In the negative-polarity ribbon within the IRIS-SG region, we find a moderate correlation: $r_S(\bar{\mathcal{D}}^-,\nu^{-}_\text{non-thermal})=0.4$. {No significant monotonic correlation is found for the positive polarity ribbon.} \cite{Cho2024OnIRIS} describes the non-thermal velocity as the cumulative effect of multi-component plasma motions (beyond that explained by local thermodynamic motions of the plasma) along the LOS direction. Although some numerical experiments show that Si IV emission can be impacted by optical-depth effects (\citealt{Kerr2019SISimulations}) it is usually characterized as optically thin plasma. Under the optically thin assumption the Si IV 1402.77 \AA{} non-thermal velocity signatures are usually interpreted as being generated by turbulence. This turbulence has been reported to be generated by the {interaction of waves (e.g. \citealt{Jeffrey2018TheFlare}) across a wide spectrum of frequencies (broadband wave interaction)} or the plasmoid instability (\citealt{French2021ProbingDynamics}). Following this interpretation, the mean degree of complexity examined in this work ($\bar{\mathcal{D}}$) may represent the POS manifestation of macroscopic effects of turbulence, while the non-thermal velocity provides the LOS proxy of the same underlying plasma dynamics. Together, these measurements could inform us of a volumetric measure of the degree of turbulence in the lower atmospheric layers. Importantly, such an interpretation may explain the moderate correlation between the reconnection rates and the mean correlation dimension: $r_S(|d\Phi/dt|^{\pm},\bar{\mathcal{D}}^{\pm}) = \{0.4,0.3\}$ (SJI observations in Figure~\ref{fig:res_CDM_hist}). Footpoints of newly reconnected field lines could be heated and swirled continuously due to turbulent motions and dissipation over timescales that differ from the larger bursts observed in the reconnection rates, 1330 \AA{}, and $50 - 100$ keV HXR emission, which would correspond to non-thermal particle injection episodes. 

Other potential contributions to the lack of strong correlation between the CDM complexity metric and indicators of reconnection dynamics are discussed bellow. First, the reconnection flux rate measure must be modified following the results of \cite{Wyper2015QuantifyingLayers}, to ensure that the evolution of the substructure complexity can be linked to the reconnection dynamics associated with current sheet plasmoid, which is outside the scope of this project. Additionally, one could also expect mechanisms such as the apparent slipping motion of magnetic fields within a quasi-separatrix layer \citep{Priest1992MagneticPoints,Priest1995ThreedimensionalFlipping} to be relevant in the evolution of the flare ribbon intensity substructure \citep{Li2015,Purkhart2025SpatiotemporalAcceleration}. Additionally, slipping reconnection has also been associated with Si IV non-thermal velocity enhancements \citep{Lorincik2022RapidCadence}, and flare ribbon co-spatial HXR sources \citep{Purkhart2025SpatiotemporalAcceleration} in previous observations. Another feasible contributor to the complexity of flare ribbon emission is the transmission of trapped non-thermal particles and heat fronts from the above-the-loop-top (ALT) region into the lower atmospheric layers. Recent studies have shown that the ALT region can host a {variety of dynamical processes} including multiple termination shocks \citep{Shibata2023NumericalFlare,Chen2015ParticleShock, French2024DopplerFlare}, plasma instabilities \citep{Shen2022TheFlares}, turbulence \citep{Shen2023Non-thermalEruptions, Ashfield2024NonthermalAcceleration, Bacchini2024ParticleLoops,  Xie2025AnisotropicFlares, French2024X-RayFan/Looptop}, magnetic traps \citep{Chen2024EnergeticArcade}, and secondary reconnection episodes \citep{Koomanski2007TheKernel,Milligan2010EVIDENCEEVENT,Kong2020DynamicalRegion}. All of these processes could contribute to the continual energy gains of charged particles. {Recently, a 3D flare model has shown that electrons escaping the ALT can generate HXR signatures in flare ribbons \citep{Li2025EnergyFlares}. Our results do not allow us to distinguish between these relative contributions to the flare ribbon substructure evolution.}  

Altogether, our analysis is a first step towards  quantitative linking of flare ribbon morphology to reconnection dynamics. We find that {the box-counting dimension appears to provide a useful observational diagnostic of multi-scale reconnection behavior, while the CDM metric offers complementary insight into localized ribbon structure.} Higher-resolution observations with instruments like the Daniel K. Inouye Solar Telescope (DKIST; \citealt{Rimmele2020TheOverview}), and coordinated on-disk and off-limb observations of flares with {the} Solar Orbiter (\citealt{Muller2020TheMission}), will provide further evidence to understand the exact physical mechanism(s) dominating the evolution of flare ribbon substructures. Other advances in flare modeling like kglobal \citep{Drake2019ASystems,Arnold2019Large-scaleModel,Drake2025MagneticWind} will allow for a more self-consistent development of non-thermal particle populations during flares at macro-scales (e.g. plasmoids). Additionally, MPI-AMRVAC \citep{Keppens2012ParallelMagnetohydrodynamics,Druett2023ChromosphericModels,Druett2024ExploringMPI-AMRVAC} flare experiments have made significant advances to the self-consistent treatment of multi-dimensional energy transport during flares and its impact on the lower solar atmosphere. These advances would be crucial to understanding the diagnostic potential of multi-scale, temporally- and spatially-variable flare ribbon observations.

\section{Conclusions}\label{sec:con}

 In this paper, we introduce a new method for measuring the multi-spatial scale flare ribbon structure based on object extraction techniques, box-counting dimension, and correlation dimension mapping. We first test our methodology on synthetic datasets (Appendix~\ref{app:FRBLE} and \ref{app:CDM}). We then apply the methodology to impulsive phase high-resolution and {high-cadence} IRIS observations of an M-class flare. Our findings are summarized as {follows}:
\begin{itemize}
    \item We find a strong statistical correlation and power law relationship between the reconnection flux rate and the box-counting dimension of the flare ribbon. This result suggests that reconnection layers are spatially fragmented and dynamical \citep{Forbes2000WhatEjections}. These results are consistent with earlier observations (e.g.   \citealt{Nakariakov2009,Cannon2023Magnetic22,Collier2024LocalisingFlare,CorchadoAlbelo2024InferringRates,Purkhart2025SpatiotemporalAcceleration}) and 3D flare simulations (e.g. \citealt{Wyper2021IsSheet, Wang2023Three-dimensionalSheet, Dahlin2025DeterminingReconnection, Wang2025BasicSheets}).
    \item We find that high spatial complexity substructures ($\mathcal{D} \gtrsim 1.3$) are localized in time and space and disappear quickly (within $19 - 95$ seconds or $1 - 5$ IRIS-SJI frames) {from the observations}. Our temporal analysis of the mean FRBLE substructure complexity in each magnetic polarity shows that most of the enhancements in spatial complexity (bursts traced by the $95^{th}$ percentile of the histogram in Figure~\ref{fig:res_CDM_hist}) are moderately statistically correlated with the reconnection flux rate (in both polarity ribbons) and the non-thermal velocities (in negative polarity ribbon) derived from the Si IV 1402.77 \AA{}. These results suggest that the correlation dimension at small scales (measure of how deformed substructures are) is not directly linked to the global reconnection properties (e.g. reconnection flux rate, non-thermal velocities). Rather, in the context of this study, CDM is most useful for spatially identifying substructures that contribute to enhanced values of the box-counting dimension.
    Our results do not allow us to distinguish the physical origin of the evolution of the flare ribbon substructure (e.g., plasmoid evolution, slipping motions, ALT region transport, or turbulent motions in the flaring atmosphere).
    \item In summary, we find strong evidence to suggest that magnetic reconnection occurs in a fragmented current sheet during the impulsive phase of this M-class flare observed on 2015-06-22 \citep{Wyper2021IsSheet, Guidoni2022SpectralIslands, Dahlin2025DeterminingReconnection, Wang2025BasicSheets}. {Our analysis can be applied to other high-resolution flare ribbon observations to calculate the commonality of the fragmented current sheet scenario, and further validate the methodology. Additional tests of simulations are required to draw physical conclusions about the evolution of the smallest substructures characterized by the CDM procedure.}
\end{itemize}

We acknowledge support from NASA ECIP NNH18ZDA001N, and NSF CAREER SPVKK1RC2MZ3 (MDK, MFCA). MFCA thanks Caroline Evans for the discussion of the Leibniz integral formalism. MFCA also thanks Dr. Judith T. Karpen for valuable discussion and manuscript revision. Support for this work is provided by the National Science Foundation through the DKIST Ambassadors program, administered by the National Solar Observatory and the Association of Universities for Research in Astronomy, Inc. RJF thanks support from the NASA HGI award 80NSSC25K7927. CAT acknowledges support from NSF grant number 2407850. VMU was partly supported by the SESDA subcontract agreement administered by ADNET.

\appendix
In this section, we first explore the physical implication of the power law relationship observed between $d\Phi/dt$ and $\mathcal{D}_\text{BC}$, show 2 examples of our methodology applied to synthetic data, and later review the expected improvements to these methods when using higher-resolution and {high-cadence} observations.
\\
\section{Evidence of the Multi-Scale Current Sheet}\label{app:MultiScaleArg}
As described in Section~\ref{sec:intr}, Equation~\ref{eq:flux} allows us to relate the  magnetic reconnection flux rate, observed in the lower solar atmosphere, to the curl of the non-ideal coronal electric field ($\nabla \times \textbf{R}^{\text{cor}})_{\parallel}$ which describes the reconnection physics. In our study, we have found that there is a power law relationship between the reconnection flux rate and the box-counting dimension (see Figure~\ref{fig:log-log_corr}). The functional form can be written as $d\Phi/dt(N,\varepsilon,t) = m\mathcal{D}_\text{BC}^{\gamma}(N,\varepsilon,t)$, where $m$ and $\gamma$ are constants determined from power law fits, and the number $N$ of boxes with spatial scale $\varepsilon$ is used to describe the FRBLE box-counting dimension. The coronal area of integration in Equation~\ref{eq:flux} is arbitrary if it {includes} all of the non-ideal electric field terms related to the change of magnetic flux. We consider the case in which the enclosing area is a smooth constant flux surface in the corona that includes all the terms $(\nabla \times \textbf{R}^{\text{cor}})_{\parallel}$. By substituting the power law relationship with the {left-hand side of the} Equation~\ref{eq:flux} we find that
\begin{align}
  m\mathcal{D}_\text{BC}^\gamma(N,\varepsilon,t) & =-\int\int_A(\nabla \times \textbf{R}^{\text{cor}})_{\parallel} dA,
\end{align}
implying that $(\nabla \times \textbf{R}^{\text{cor}})_{\parallel} = \mathcal{F}(N,\varepsilon,t)$. This relationship suggests that the physics involved in the reconnection process must be described by a multi-spatial-scale function $\mathcal{F}(N,\varepsilon,t)$. Such {multi-spatial-scale} reconnection physics supports the fragmented or inhomogeneous current sheet scenarios (e.g. {plasmoid-unstable or 3D-turbulent} current sheets).

\section{FRBLE Procedure on Synthetic Data}\label{app:FRBLE}
\begin{figure}
    \centering
    \includegraphics[width=\linewidth]{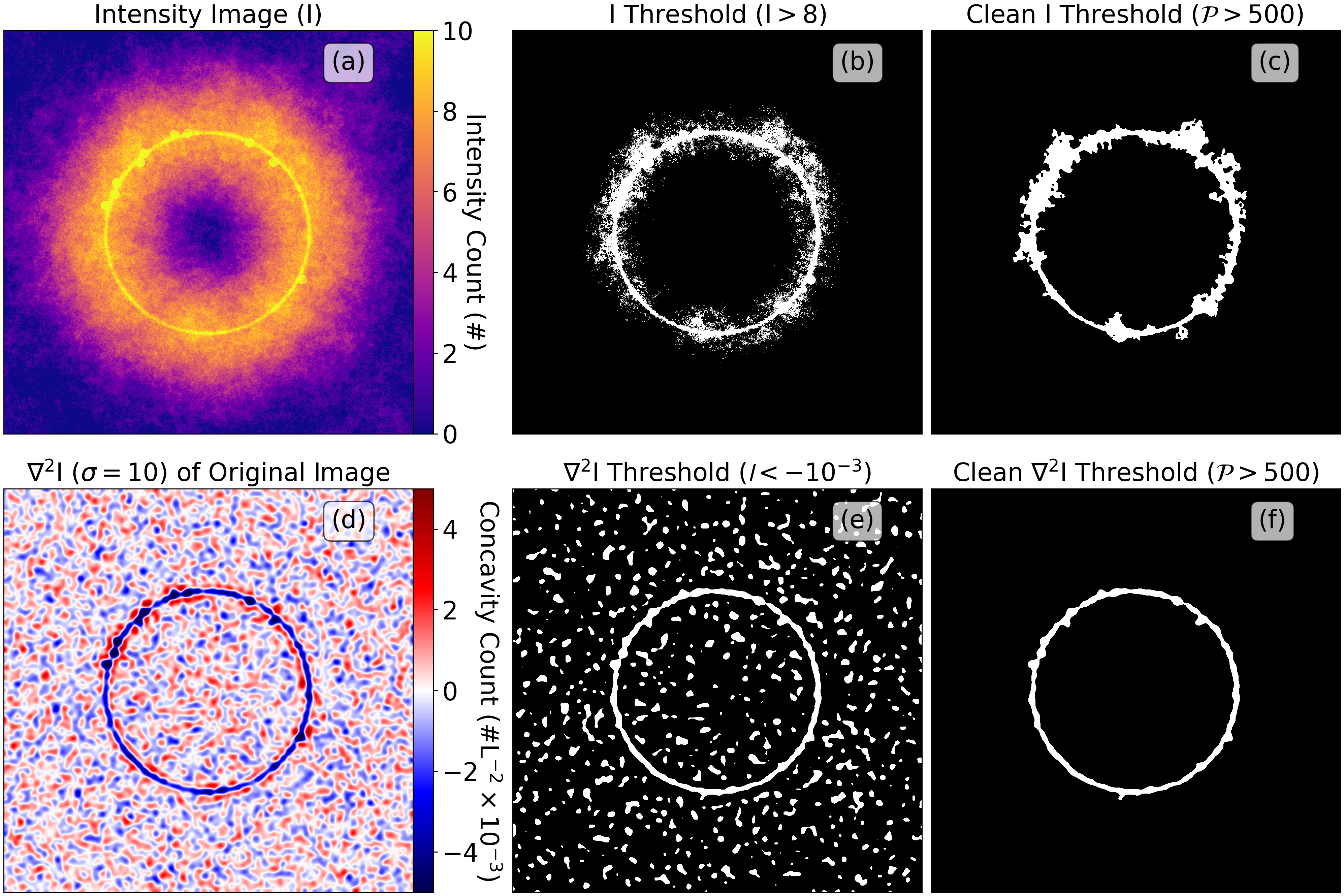}
    \caption{{FRBLE extraction test on synthetic data. (a) Synthetic image: a ring with embedded circular substructures plus noise. (b,c) features extracted using a global intensity threshold, before and after cleaning. (d) Intensity curvature ($\nabla^2$I) map. (e,f) Features extracted using a curvature threshold, before and after cleaning.}}
    \label{fig:met_FRBLE}
\end{figure}

A test of the object extraction technique procedure (FRBLE) is shown in Figure~\ref{fig:met_FRBLE}. We start by creating a synthetic ring structure and place circular substructures around the circumference. We then add white and colored noise ($\text{signal} \approx f^\beta$) to the image so that it resembles solar observations. The top row shows how a high intensity threshold ($I \geq 10$ DN) is used to extract the ring and circle substructures. The rightmost column shows the extracted features, after applying the cleaning algorithm. The binary image (c) shows that the extraction based on a global intensity threshold overestimates the features of interest (ring and circle substructures). The overestimation of the features is likely due to the high level of noise contamination and low contrast between background and features. The bottom row shows the application of the threshold based on the intensity curvature of the image ($\nabla^2 I$). We approximate the intensity curvature using a LoG operator with a $\sigma=10$ pixel for noise reduction in the spatial derivative calculations. {The final binary image (f) shows that the substructure has been successfully extracted, after applying the curvature threshold $\nabla^2 I \leq - 1\times 10^{-3}$ DN/pixel$^2$ and the cleaning procedure. }

\section{Correlation Dimension Mapping on Synthetic Multi-Scale Boundary Object}\label{app:CDM}
\begin{figure}
    \centering
    \includegraphics[width=\linewidth]{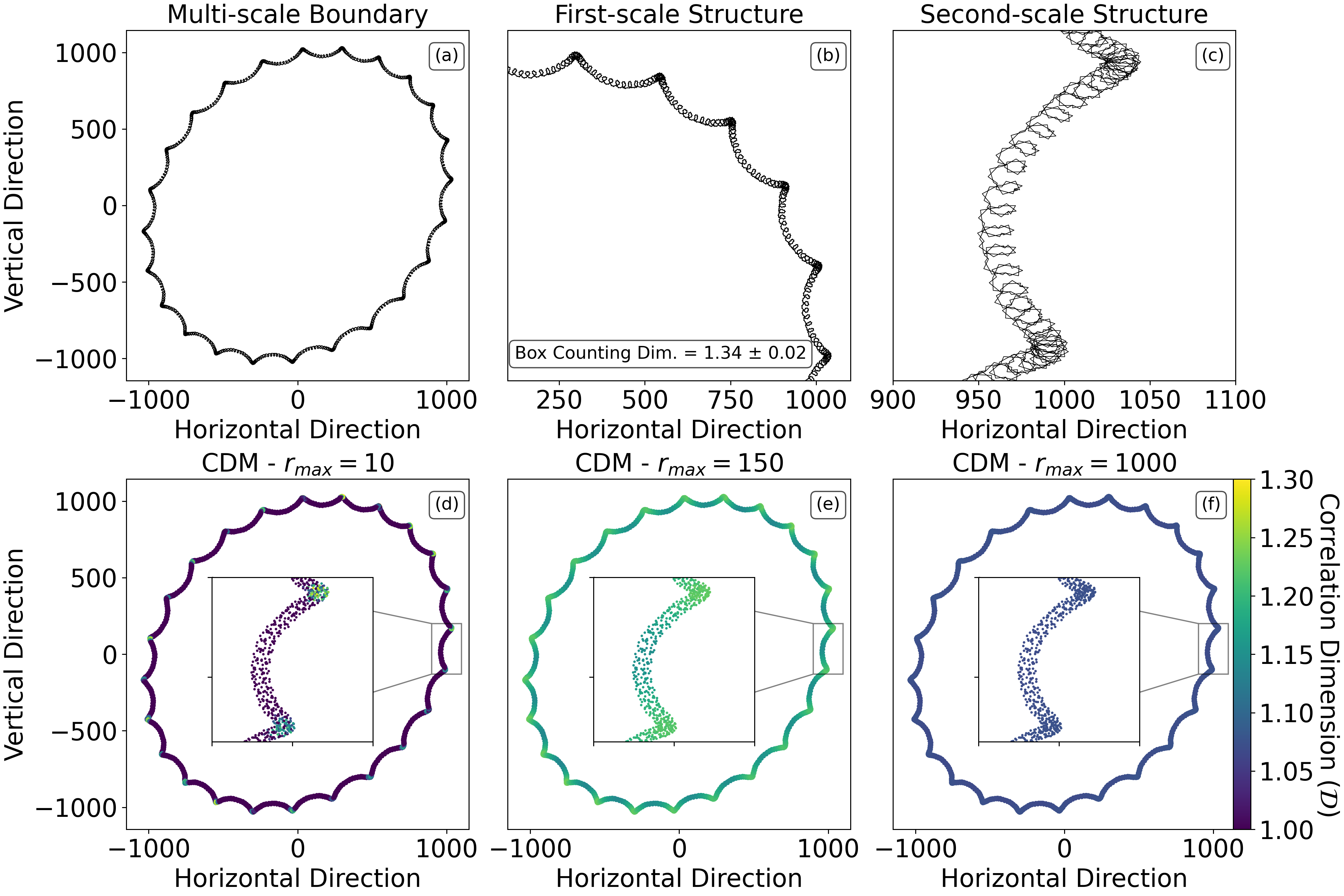}
    \caption{{CDM on a synthetic multi-scale boundary and sensitivity to $r_{\rm max}$. (a--c) The boundary with multiple embedded spatial scales. (d-f) CDM results for $r_{\rm max}=\{10,150,1000\}$, illustrating how different scale ranges control the inferred $\mathcal{D}$ values and identification of high-complexity regions}}
    \label{fig:met_CDM_r}
\end{figure}

Another test applied to synthetic data shows (Figure~\ref{fig:met_CDM_r}) the implementation of the CDM algorithm. We define a multi-scale boundary, top row of the figure, using a combination of sinusoidal functions and Weierstrass functions (\citealt{Weierstra1988UberBesitzen}). The x and y coordinates of the boundary are defined as: $x = 1000\cos(\pi t) + \sum_{\text{i}=0}^5 0.3^i  \sin(23^i \pi t)$ and $y = 1000\sin(\pi t) + \sum_{\text{i}=0}^5 0.3^i  \cos(23^i \pi t)$. The box-counting dimension of this boundary is $\mathcal{D}_\text{BC} = 1.37 \pm 0.02$, which describes the multi-scale and space filling properties of the synthetic boundary. The bottom row highlights how the choice of $r_{\text{max}}$ in the CDM procedure determines the correlation dimension value $\mathcal{D}$ and the location of high complexity regions. Overall, in regions where the boundary kinks the correlation dimension is higher, due to the substructure's high spatial complexity (panels d and e). If the $r_{\text{max}}$ is comparable to the size of the main structure, CDM shows a more uniform spatial distribution and lower values of $\mathcal{D}$ (panel e). However, when tuned to spatial scales of the same order of magnitude as the kinked or folded regions, the CDM map highlights these regions of varying complexity through higher localized values of $\mathcal{D}$ (panel e). When using CDM, scales of known interest can be input \textit{a priori}, or a series of tests with varying $r$ ranges can be tested to extract the most relevant spatial scales for the data.
\section{Expected Improvements with Higher-Resolution and Higher-Cadence Data}\label{app:improv}
It should be noted that the current implementation of the FRBLE tracking algorithm and CDM characterization of the substructure complexity are mainly dependent on two qualities of the dataset -- the contrast between the FRBLE and the rest of the emitting atmosphere, and the spatial resolution of the observations. We employ spatial derivative filters on the IRIS intensity maps to circumvent the contrast between the FRBLE and the rest of the flare ribbon emission, which is extracted using intensity thresholds \citep{Qiu2004MagneticEvents,Maurya2010ARelease,Kirk2013AnBrightenings,Yang2018AutomatedImages}. Yet, due to the sensitivity of the spatial derivatives to noise and pixel intensity saturation, we know that higher cadence (e.g. IRIS high-cadence program, Hi-C, MUSE, and SO-EUI) and higher spatial resolution (e.g. DKIST, GST, and SST) observations can provide a more accurate extraction of the flare ribbon substructure. Currently, to reduce potential artifacts (small scale morphology changes due to noise in the intensity maps) in the FRBLE extraction, we smooth the extracted boundary objects using a {low-pass filter}. Doing so reduces the spatial scales used to characterize the FRBLE substructures with CDM between 2.4 - 3.6 Mm (corresponding to local radii of 10 - 15 pixels of the IRIS-SJI camera, with 0.33 arcsecond resolution, at each boundary coordinate). DKIST Visible Broadband Imager (VBI; \citealt{Woger2021TheVBI}) observations have an expected spatial resolution of 0.017 arcseconds \citep{Tamburri2025UnveilingDKIST}; thus, even when losing the same 10 pixels in the extraction of the FRBLE, it will still improve the lower bound of the spatial scales used in CDM to approximately 0.7 Mm. Furthermore, the spatial resolution of DKIST-VBI will improve the sampling of the spatial scales 20 times the current implementation, meaning DKIST-VBI can observe approximately 20 pixels per each IRIS-SJI pixel.

\bibliographystyle{aasjournal}
\bibliography{references}

@article{Kazachenko2017,
    title = {{A Database of Flare Ribbon Properties from the Solar Dynamics Observatory . I. Reconnection Flux}},
    year = {2017},
    journal = {The Astrophysical Journal},
    author = {Kazachenko, Maria D. and Lynch, Benjamin J. and Welsch, Brian T. and Sun, Xudong},
    number = {1},
    month = {8},
    pages = {49},
    volume = {845},
    publisher = {IOP Publishing},
    url = {http://dx.doi.org/10.3847/1538-4357/aa7ed6 https://iopscience.iop.org/article/10.3847/1538-4357/aa7ed6},
    doi = {10.3847/1538-4357/aa7ed6},
    issn = {1538-4357},
    arxivId = {1704.05097}
}

@article{Forbes2000,
    title = {{A review on the genesis of coronal mass ejections}},
    year = {2000},
    journal = {Journal of Geophysical Research: Space Physics},
    author = {Forbes, T. G.},
    number = {A10},
    pages = {23153--23165},
    volume = {105},
    url = {http://doi.wiley.com/10.1029/2000JA000005},
    isbn = {0148-0227},
    doi = {10.1029/2000JA000005},
    issn = {01480227}
}

@article{Fletcher2008,
    title = {{Impulsive Phase Flare Energy Transport by Large‐Scale Alfv{\'{e}}n Waves and the Electron Acceleration Problem}},
    year = {2008},
    journal = {The Astrophysical Journal},
    author = {Fletcher, L. and Hudson, H. S.},
    number = {2},
    month = {3},
    pages = {1645--1655},
    volume = {675},
    url = {https://iopscience.iop.org/article/10.1086/527044},
    doi = {10.1086/527044},
    issn = {0004-637X}
}

@article{Hayes2019,
    title = {{Persistent Quasi-Periodic Pulsations During a Large X-Class Solar Flare}},
    year = {2019},
    journal = {arXiv},
    author = {Hayes, Laura A. and Gallagher, Peter T. and Dennis, Brian R. and Ireland, Jack and Inglis, Andrew and Morosan, Diana E.},
    number = {1},
    pages = {33},
    volume = {875},
    publisher = {IOP Publishing},
    url = {http://dx.doi.org/10.3847/1538-4357/ab0ca3},
    doi = {10.3847/1538-4357/ab0ca3},
    issn = {23318422},
    arxivId = {1903.01328}
}

@article{Nakariakov2009,
    title = {{Quasi-periodic pulsations in solar flares}},
    year = {2009},
    journal = {Space Science Reviews},
    author = {Nakariakov, V. M. and Melnikov, V. F.},
    number = {1-4},
    pages = {119--151},
    volume = {149},
    doi = {10.1007/s11214-009-9536-3},
    issn = {00386308}
}

@article{Li2015,
    title = {{QUASI-PERIODIC SLIPPING MAGNETIC RECONNECTION DURING AN X-CLASS SOLAR FLARE OBSERVED BY THE SOLAR DYNAMICS OBSERVATORY AND INTERFACE REGION IMAGING SPECTROGRAPH}},
    year = {2015},
    journal = {The Astrophysical Journal},
    author = {Li, Ting and Zhang, Jun},
    number = {1},
    month = {4},
    pages = {L8},
    volume = {804},
    publisher = {IOP Publishing},
    url = {http://dx.doi.org/10.1088/2041-8205/804/1/L8 https://iopscience.iop.org/article/10.1088/2041-8205/804/1/L8},
    doi = {10.1088/2041-8205/804/1/L8},
    issn = {2041-8213}
}

@article{Drake2019ASystems,
    title = {{A computational model for exploring particle acceleration during reconnection in macroscale systems}},
    year = {2019},
    journal = {Physics of Plasmas},
    author = {Drake, J. F. and Arnold, H. and Swisdak, M. and Dahlin, J. T.},
    number = {1},
    month = {1},
    volume = {26},
    doi = {10.1063/1.5058140},
    issn = {1070-664X}
}

@article{Fan2024A11158,
    title = {{A Data-driven Magnetohydrodynamic Simulation of the 2011 February 15 Coronal Mass Ejection from Active Region NOAA 11158}},
    year = {2024},
    journal = {The Astrophysical Journal},
    author = {Fan, Yuhong and Kazachenko, Maria D. and Afanasyev, Andrey N. and Fisher, George H.},
    number = {2},
    month = {11},
    pages = {206},
    volume = {975},
    doi = {10.3847/1538-4357/ad7f53},
    issn = {0004-637X}
}

@article{Drake2010ARAYS,
    title = {{A MAGNETIC RECONNECTION MECHANISM FOR THE GENERATION OF ANOMALOUS COSMIC RAYS}},
    year = {2010},
    journal = {The Astrophysical Journal},
    author = {Drake, J. F. and Opher, M. and Swisdak, M. and Chamoun, J. N.},
    number = {2},
    month = {2},
    pages = {963--974},
    volume = {709},
    doi = {10.1088/0004-637X/709/2/963},
    issn = {0004-637X}
}

@article{DePontieu2021AIRIS,
    title = {{A New View of the Solar Interface Region from the Interface Region Imaging Spectrograph (IRIS)}},
    year = {2021},
    journal = {Solar Physics},
    author = {De Pontieu, Bart and Polito, Vanessa and Hansteen, Viggo and Testa, Paola and Reeves, Katharine K. and Antolin, Patrick and N{\'{o}}brega-Siverio, Daniel Elias and Kowalski, Adam F. and Martinez-Sykora, Juan and Carlsson, Mats and McIntosh, Scott W. and Liu, Wei and Daw, Adrian and Kankelborg, Charles C.},
    number = {5},
    month = {5},
    pages = {84},
    volume = {296},
    publisher = {Springer Science and Business Media B.V.},
    url = {https://link.springer.com/10.1007/s11207-021-01826-0},
    doi = {10.1007/s11207-021-01826-0},
    issn = {0038-0938}
}

@inproceedings{Carmichael1964AFlares,
    title = {{A Process for Flares}},
    year = {1964},
    booktitle = {The Physics of Solar Flares},
    author = {Carmichael, H.},
    pages = {451--456},
    url = {https://ui.adsabs.harvard.edu/abs/1964NASSP..50..451C/abstract}
}

@article{Longcope2014AFLARE,
    title = {{A SIMPLE MODEL OF CHROMOSPHERIC EVAPORATION AND CONDENSATION DRIVEN CONDUCTIVELY IN A SOLAR FLARE}},
    year = {2014},
    journal = {The Astrophysical Journal},
    author = {Longcope, D. W.},
    number = {1},
    month = {10},
    pages = {10},
    volume = {795},
    doi = {10.1088/0004-637X/795/1/10},
    issn = {1538-4357}
}

@article{Maurya2010ARelease,
    title = {{A Technique for Automated Determination of Flare Ribbon Separation and Energy Release}},
    year = {2010},
    journal = {Solar Physics},
    author = {Maurya, R. A. and Ambastha, A.},
    number = {2},
    month = {4},
    pages = {337--353},
    volume = {262},
    doi = {10.1007/s11207-009-9488-5},
    issn = {0038-0938}
}

@article{Hesse1988AReconnection,
    title = {{A theoretical foundation of general magnetic reconnection}},
    year = {1988},
    journal = {Journal of Geophysical Research: Space Physics},
    author = {Hesse, M. and Schindler, K.},
    number = {A6},
    month = {6},
    pages = {5559--5567},
    volume = {93},
    doi = {10.1029/JA093iA06p05559},
    issn = {0148-0227}
}

@article{Reep2016ALFVENICFLARES,
    title = {{ALFV{\'{E}}NIC WAVE HEATING OF THE UPPER CHROMOSPHERE IN FLARES}},
    year = {2016},
    journal = {The Astrophysical Journal Letters},
    author = {Reep, J. W. and Russell, A. J. B.},
    number = {1},
    month = {2},
    pages = {L20},
    volume = {818},
    doi = {10.3847/2041-8205/818/1/L20},
    issn = {2041-8205}
}

@article{Kirk2013AnBrightenings,
    title = {{An Automated Algorithm to Distinguish and Characterize Solar Flares and Associated Sequential Chromospheric Brightenings}},
    year = {2013},
    journal = {Solar Physics},
    author = {Kirk, M. S. and Balasubramaniam, K. S. and Jackiewicz, J. and McNamara, B. J. and McAteer, R. T. J.},
    number = {1},
    month = {3},
    pages = {97--111},
    volume = {283},
    doi = {10.1007/s11207-011-9843-1},
    issn = {0038-0938}
}

@article{Xie2025AnisotropicFlares,
    title = {{Anisotropic Turbulent Flows Observed in Above-the-loop-top Regions during Solar Flares}},
    year = {2025},
    journal = {The Astrophysical Journal Letters},
    author = {Xie, Xiaoyan and Shen, Chengcai and Reeves, Katharine K. and Chen, Bin and Li, Xiaocan and Guo, Fan and Yu, Sijie and Wei, Yuqian and Dong, Chuanfei},
    number = {1},
    month = {5},
    pages = {L27},
    volume = {984},
    doi = {10.3847/2041-8213/adc91b},
    issn = {2041-8205}
}

@article{Huang2022AuroralRegion,
    title = {{Auroral Spiral Structure Formation Through Magnetic Reconnection in the Auroral Acceleration Region}},
    year = {2022},
    journal = {Geophysical Research Letters},
    author = {Huang, Kai and Liu, Yi‐Hsin and Lu, Quanming and Hu, Zejun and Lynch, Kristina A. and Hesse, Michael and Vaivads, Andris and Yang, Huigen},
    number = {18},
    month = {9},
    volume = {49},
    doi = {10.1029/2022GL100466},
    issn = {0094-8276}
}

@article{Yang2018AutomatedImages,
    title = {{Automated Solar Flare Detection and Feature Extraction in High-Resolution and Full-Disk H{\$}{\textbackslash}upalpha{\$} Images}},
    year = {2018},
    journal = {Solar Physics},
    author = {Yang, Meng and Tian, Yu and Liu, Yangyi and Rao, Changhui},
    number = {5},
    month = {5},
    pages = {81},
    volume = {293},
    doi = {10.1007/s11207-018-1300-y},
    issn = {0038-0938}
}

@article{Qu2003AutomaticSVM,
    title = {{Automatic Solar Flare Detection Using MLP, RBF, and SVM}},
    year = {2003},
    journal = {Solar Physics},
    author = {Qu, Ming and Shih, Frank Y. and Jing, Ju and Wang, Haimin},
    number = {1},
    month = {10},
    pages = {157--172},
    volume = {217},
    doi = {10.1023/A:1027388729489},
    issn = {0038-0938}
}

@article{Wang2025BasicSheets,
    title = {{Basic Pattern of Three-dimensional Magnetic Reconnection within Strongly Turbulent Current Sheets}},
    year = {2025},
    journal = {The Astrophysical Journal},
    author = {Wang, Yulei and Cheng, Xin and Ding, Mingde},
    number = {1},
    month = {5},
    pages = {43},
    volume = {985},
    doi = {10.3847/1538-4357/adca3d},
    issn = {0004-637X}
}

@article{Ripperda2022BlackReconnection,
    title = {{Black Hole Flares: Ejection of Accreted Magnetic Flux through 3D Plasmoid-mediated Reconnection}},
    year = {2022},
    journal = {The Astrophysical Journal Letters},
    author = {Ripperda, B. and Liska, M. and Chatterjee, K. and Musoke, G. and Philippov, A. A. and Markoff, S. B. and Tchekhovskoy, A. and Younsi, Z.},
    number = {2},
    month = {1},
    pages = {L32},
    volume = {924},
    doi = {10.3847/2041-8213/ac46a1},
    issn = {2041-8205}
}

@article{Nyquist1928CertainTheory,
    title = {{Certain Topics in Telegraph Transmission Theory}},
    year = {1928},
    journal = {Transactions of the American Institute of Electrical Engineers},
    author = {Nyquist, H.},
    number = {2},
    month = {4},
    pages = {617--644},
    volume = {47},
    doi = {10.1109/T-AIEE.1928.5055024},
    issn = {0096-3860}
}

@article{Druett2023ChromosphericModels,
    title = {{Chromospheric Evaporation by Particle Beams in Multi-Dimensional Flare Models}},
    year = {2023},
    journal = {Solar Physics},
    author = {Druett, Malcolm Keith and Ruan, Wenzhi and Keppens, Rony},
    number = {11},
    month = {11},
    pages = {134},
    volume = {298},
    doi = {10.1007/s11207-023-02224-4},
    issn = {0038-0938}
}

@article{Kurokawa1988CloseFlare,
    title = {{Close Relationship between H{$\alpha$} and Hard X-Ray Emissions at the Impulsive Phase of a Solar Flare}},
    year = {1988},
    journal = {Publications of the Astronomical Society of Japan},
    author = {Kurokawa, Hiroki and Takakura, Tatsuo and Ohki, Kenichiro},
    number = {3},
    month = {7},
    pages = {357--367},
    volume = {40},
    doi = {10.1093/pasj/40.3.357},
    issn = {2053-051X}
}

@article{Shannon1949CommunicationNoise,
    title = {{Communication in the Presence of Noise}},
    year = {1949},
    journal = {Proceedings of the IRE},
    author = {Shannon, C.E.},
    number = {1},
    month = {1},
    pages = {10--21},
    volume = {37},
    doi = {10.1109/JRPROC.1949.232969},
    issn = {0096-8390}
}

@article{Naus2022CorrelatedFlare,
    title = {{Correlated Spatio-temporal Evolution of Extreme-Ultraviolet Ribbons and Hard X-Rays in a Solar Flare}},
    year = {2022},
    journal = {The Astrophysical Journal},
    author = {Naus, S. J. and Qiu, J. and DeVore, C. R. and Antiochos, S. K. and Dahlin, J. T. and Drake, J. F. and Swisdak, M.},
    number = {2},
    month = {2},
    pages = {218},
    volume = {926},
    publisher = {American Astronomical Society},
    doi = {10.3847/1538-4357/ac4028},
    issn = {0004-637X},
    arxivId = {2109.15314}
}

@article{Dahlin2025DeterminingReconnection,
    title = {{Determining the 3D Dynamics of Solar Flare Magnetic Reconnection}},
    year = {2025},
    journal = {The Astrophysical Journal},
    author = {Dahlin, Joel T. and Antiochos, Spiro K. and Wyper, Peter F. and Qiu, Jiong and DeVore, C. Richard},
    number = {1},
    month = {11},
    pages = {31},
    volume = {993},
    url = {https://iopscience.iop.org/article/10.3847/1538-4357/ae03c5},
    doi = {10.3847/1538-4357/ae03c5},
    issn = {0004-637X}
}

@article{Huang2017DevelopmentIsland,
    title = {{Development of Turbulent Magnetic Reconnection in a Magnetic Island}},
    year = {2017},
    journal = {The Astrophysical Journal},
    author = {Huang, Can and Lu, Quanming and Wang, Rongsheng and Guo, Fan and Wu, Mingyu and Lu, San and Wang, Shui},
    number = {2},
    month = {2},
    pages = {245},
    volume = {835},
    doi = {10.3847/1538-4357/835/2/245},
    issn = {0004-637X}
}

@article{Huang2012DistributionReconnection,
    title = {{Distribution of Plasmoids in High-Lundquist-Number Magnetic Reconnection}},
    year = {2012},
    journal = {Physical Review Letters},
    author = {Huang, Yi-Min and Bhattacharjee, A.},
    number = {26},
    month = {12},
    pages = {265002},
    volume = {109},
    doi = {10.1103/PhysRevLett.109.265002},
    issn = {0031-9007}
}

@article{French2024DopplerFlare,
    title = {{Doppler signature of a possible termination shock in an off-limb solar flare}},
    year = {2024},
    journal = {Monthly Notices of the Royal Astronomical Society},
    author = {French, Ryan J and Yu, Sijie and Chen, Bin and Shen, Chengcai and Matthews, Sarah A},
    number = {4},
    month = {2},
    pages = {6836--6844},
    volume = {528},
    doi = {10.1093/mnras/stae430},
    issn = {0035-8711}
}

@article{French2025DualFlareb,
    title = {{Dual Origins of Rapid Flare Ribbon Downflows in an X9-class Solar Flare}},
    year = {2025},
    journal = {The Astrophysical Journal},
    author = {French, Ryan J. and Ashfield, William H. and Tamburri, Cole A. and Kazachenko, Maria D. and Dominique, Marie and Albelo, Marcel Corchado and Caspi, Amir},
    number = {2},
    month = {12},
    pages = {182},
    volume = {995},
    doi = {10.3847/1538-4357/ae21b5},
    issn = {0004-637X}
}

@article{Kong2020DynamicalRegion,
    title = {{Dynamical Modulation of Solar Flare Electron Acceleration due to Plasmoid-shock Interactions in the Looptop Region}},
    year = {2020},
    journal = {The Astrophysical Journal Letters},
    author = {Kong, Xiangliang and Guo, Fan and Shen, Chengcai and Chen, Bin and Chen, Yao and Giacalone, Joe},
    number = {2},
    month = {12},
    pages = {L16},
    volume = {905},
    doi = {10.3847/2041-8213/abcbf5},
    issn = {2041-8205}
}

@book{Heaviside2011ElectricalPapers,
    title = {{Electrical Papers}},
    year = {2011},
    author = {Heaviside, Oliver},
    month = {6},
    publisher = {Cambridge University Press},
    url = {https://www.cambridge.org/core/product/identifier/9780511983139/type/book},
    isbn = {9780511983139},
    doi = {10.1017/CBO9780511983139}
}

@article{Drake2006ElectronReconnection,
    title = {{Electron acceleration from contracting magnetic islands during reconnection}},
    year = {2006},
    journal = {Nature},
    author = {Drake, J. F. and Swisdak, M. and Che, H. and Shay, M. A.},
    number = {7111},
    month = {10},
    pages = {553--556},
    volume = {443},
    publisher = {Nature Publishing Group},
    doi = {10.1038/nature05116},
    issn = {14764687}
}

@article{Chen2024EnergeticArcade,
    title = {{Energetic Electrons Accelerated and Trapped in a Magnetic Bottle above a Solar Flare Arcade}},
    year = {2024},
    journal = {The Astrophysical Journal},
    author = {Chen, Bin and Kong, Xiangliang and Yu, Sijie and Shen, Chengcai and Li, Xiaocan and Guo, Fan and Zhang, Yixian and Glesener, Lindsay and Krucker, Säm},
    number = {1},
    month = {8},
    pages = {85},
    volume = {971},
    doi = {10.3847/1538-4357/ad531a},
    issn = {0004-637X}
}

@article{Li2025EnergyFlares,
    title = {{Energy Conversion and Electron Acceleration and Transport in 3D Simulations of Solar Flares}},
    year = {2025},
    journal = {The Astrophysical Journal},
    author = {Li, Xiaocan and Shen, Chengcai and Xie, Xiaoyan and Guo, Fan and Chen, Bin and Oparin, Ivan and Wei, Yuqian and Yu, Sijie and Seo, Jeongbhin},
    number = {2},
    month = {10},
    pages = {202},
    volume = {991},
    doi = {10.3847/1538-4357/adfcd5},
    issn = {0004-637X}
}

@article{Milligan2010EVIDENCEEVENT,
    title = {{EVIDENCE OF A PLASMOID-LOOPTOP INTERACTION AND MAGNETIC INFLOWS DURING A SOLAR FLARE/CORONAL MASS EJECTION ERUPTIVE EVENT}},
    year = {2010},
    journal = {The Astrophysical Journal},
    author = {Milligan, Ryan O. and McAteer, R. T. James and Dennis, Brian R. and Young, C. Alex},
    number = {2},
    month = {4},
    pages = {1292--1300},
    volume = {713},
    doi = {10.1088/0004-637X/713/2/1292},
    issn = {0004-637X}
}

@article{French2025EvolutionFlare,
    title = {{Evolution of Flare Ribbon Bead-like Structures in a Solar Flare}},
    year = {2025},
    journal = {The Astrophysical Journal Letters},
    author = {French, Ryan J. and Kazachenko, Maria D. and Berghmans, David and D’Huys, Elke and Dominique, Marie and Patel, Ritesh and Talpeanu, Dana-Camelia and Tamburri, Cole A. and Yadav, Rahul},
    number = {2},
    month = {12},
    pages = {L54},
    volume = {995},
    url = {https://iopscience.iop.org/article/10.3847/2041-8213/ae2684},
    doi = {10.3847/2041-8213/ae2684},
    issn = {2041-8205}
}

@article{Janvier2016EvolutionFlare,
    title = {{Evolution of flare ribbons, electric currents, and quasi-separatrix layers during an X-class flare}},
    year = {2016},
    journal = {Astronomy {\&} Astrophysics},
    author = {Janvier, M. and Savcheva, A. and Pariat, E. and Tassev, S. and Millholland, S. and Bommier, V. and McCauley, P. and McKillop, S. and Dougan, F.},
    month = {7},
    pages = {A141},
    volume = {591},
    doi = {10.1051/0004-6361/201628406},
    issn = {0004-6361}
}

@article{Alfven1942ExistenceWaves,
    title = {{Existence of Electromagnetic-Hydrodynamic Waves}},
    year = {1942},
    journal = {Nature},
    author = {Alfv{\'{e}}n, H.},
    number = {3805},
    month = {10},
    pages = {405--406},
    volume = {150},
    url = {https://www.nature.com/articles/150405d0},
    doi = {10.1038/150405d0},
    issn = {0028-0836}
}

@article{Olson2016ExperimentalReconnection,
    title = {{Experimental Demonstration of the Collisionless Plasmoid Instability below the Ion Kinetic Scale during Magnetic Reconnection}},
    year = {2016},
    journal = {Physical Review Letters},
    author = {Olson, J. and Egedal, J. and Greess, S. and Myers, R. and Clark, M. and Endrizzi, D. and Flanagan, K. and Milhone, J. and Peterson, E. and Wallace, J. and Weisberg, D. and Forest, C. B.},
    number = {25},
    month = {6},
    pages = {255001},
    volume = {116},
    doi = {10.1103/PhysRevLett.116.255001},
    issn = {0031-9007}
}

@article{Druett2024ExploringMPI-AMRVAC,
    title = {{Exploring self-consistent 2.5D flare simulations with MPI-AMRVAC}},
    year = {2024},
    journal = {Astronomy {\&} Astrophysics},
    author = {Druett, Malcolm and Ruan, Wenzhi and Keppens, Rony},
    month = {4},
    pages = {A171},
    volume = {684},
    doi = {10.1051/0004-6361/202347600},
    issn = {0004-6361}
}

@article{Donnelly1971ExtremeResults,
    title = {{Extreme ultraviolet flashes of solar flares observed via sudden frequency deviations: Experimental results}},
    year = {1971},
    journal = {Solar Physics},
    author = {Donnelly, Richard F.},
    number = {1},
    month = {10},
    pages = {188--203},
    volume = {20},
    doi = {10.1007/BF00146110},
    issn = {0038-0938}
}

@article{Brosius2012EXTREME-ULTRAVIOLETFLARE,
    title = {{EXTREME-ULTRAVIOLET SPECTROSCOPIC OBSERVATION OF DIRECT CORONAL HEATING DURING A C-CLASS SOLAR FLARE}},
    year = {2012},
    journal = {The Astrophysical Journal},
    author = {Brosius, Jeffrey W.},
    number = {1},
    month = {7},
    pages = {54},
    volume = {754},
    doi = {10.1088/0004-637X/754/1/54},
    issn = {0004-637X}
}

@article{Uzdensky2010FastRegime,
    title = {{Fast magnetic reconnection in the plasmoid-dominated regime}},
    year = {2010},
    journal = {Physical Review Letters},
    author = {Uzdensky, D. A. and Loureiro, N. F. and Schekochihin, A. A.},
    number = {23},
    month = {12},
    volume = {105},
    doi = {10.1103/PhysRevLett.105.235002},
    issn = {00319007},
    arxivId = {1008.3330}
}

@article{Furth1963Finite-resistivityPinch,
    title = {{Finite-resistivity instabilities of a sheet pinch}},
    year = {1963},
    journal = {Physics of Fluids},
    author = {Furth, Harold P. and Killeen, John and Rosenbluth, Marshall N.},
    number = {4},
    pages = {459--484},
    volume = {6},
    doi = {10.1063/1.1706761},
    issn = {10706631}
}

@article{Benz2017FlareObservations,
    title = {{Flare Observations}},
    year = {2017},
    journal = {Living Reviews in Solar Physics},
    author = {Benz, Arnold O.},
    number = {1},
    month = {12},
    volume = {14},
    publisher = {Max Planck Institute for Solar System Research},
    doi = {10.1007/s41116-016-0004-3},
    issn = {16144961}
}

@article{Asai2004FlareRate,
    title = {{Flare Ribbon Expansion and Energy Release Rate}},
    year = {2004},
    journal = {The Astrophysical Journal},
    author = {Asai, Ayumi and Yokoyama, Takaaki and Shimojo, Masumi and Masuda, Satoshi and Kurokawa, Hiroki and Shibata, Kazunari},
    number = {1},
    month = {8},
    pages = {557--567},
    volume = {611},
    doi = {10.1086/422159},
    issn = {0004-637X}
}

@article{Priest2017Flux-RopeReconnection,
    title = {{Flux-Rope Twist in Eruptive Flares and CMEs: Due to Zipper and Main-Phase Reconnection}},
    year = {2017},
    journal = {Solar Physics},
    author = {Priest, E. R. and Longcope, D. W.},
    number = {1},
    month = {1},
    pages = {25},
    volume = {292},
    doi = {10.1007/s11207-016-1049-0},
    issn = {0038-0938}
}

@article{Cerutti2021FormationNebula,
    title = {{Formation of giant plasmoids at the pulsar wind termination shock: A possible origin of the inner-ring knots in the Crab Nebula}},
    year = {2021},
    journal = {Astronomy {\&} Astrophysics},
    author = {Cerutti, Benoît and Giacinti, Gwenael},
    month = {12},
    pages = {A91},
    volume = {656},
    doi = {10.1051/0004-6361/202142178},
    issn = {0004-6361}
}

@article{Schindler1988GeneralHelicity,
    title = {{General magnetic reconnection, parallel electric fields, and helicity}},
    year = {1988},
    journal = {Journal of Geophysical Research: Space Physics},
    author = {Schindler, K. and Hesse, M. and Birn, J.},
    number = {A6},
    month = {6},
    pages = {5547--5557},
    volume = {93},
    doi = {10.1029/JA093iA06p05547},
    issn = {0148-0227}
}

@article{Hudson2011GlobalFlares,
    title = {{Global Properties of Solar Flares}},
    year = {2011},
    journal = {Space Science Reviews},
    author = {Hudson, Hugh S.},
    number = {1},
    month = {1},
    pages = {5--41},
    volume = {158},
    doi = {10.1007/s11214-010-9721-4},
    issn = {0038-6308}
}

@inproceedings{Bornmann1996GOESDisturbances,
    title = {{GOES x-ray sensor and its use in predicting solar-terrestrial disturbances}},
    year = {1996},
    booktitle = {GOES-8 and Beyond},
    author = {Bornmann, Patricia L. and Speich, David and Hirman, Joseph and Matheson, Lorne and Grubb, Richard and Garcia, Howard A. and Viereck, R.},
    editor = {Washwell, Edward R.},
    month = {10},
    pages = {291--298},
    volume = {2812},
    publisher = {SPIE},
    url = {http://proceedings.spiedigitallibrary.org/proceeding.aspx?articleid=1021196},
    doi = {10.1117/12.254076}
}

@incollection{Machol2019GOES-RIrradiance,
    title = {{GOES-R Series Solar X-ray and Ultraviolet Irradiance}},
    year = {2019},
    booktitle = {The GOES-R Series: A New Generation of Geostationary Environmental Satellites},
    author = {Machol, Janet L. and Eparvier, Francis G. and Viereck, Rodney A. and Woodraska, Donald L. and Snow, Martin and Thiemann, Ed and Woods, Thomas N. and McClintock, William E. and Mueller, Steven and Eden, Thomas D. and Meisner, Randle and Codrescu, Stefan and Bouwer, S. Dave and Reinard, Alysha A.},
    month = {10},
    pages = {233--242},
    publisher = {Elsevier},
    isbn = {9780128143285},
    doi = {10.1016/B978-0-12-814327-8.00019-6}
}

@article{Huang2014HEruption,
    title = {{H <i>{$\alpha$}</i> spectroscopy and multiwavelength imaging of a solar flare caused by filament eruption}},
    year = {2014},
    journal = {Astronomy {\&} Astrophysics},
    author = {Huang, Z. and Madjarska, M. S. and Koleva, K. and Doyle, J. G. and Duchlev, P. and Dechev, M. and Reardon, K.},
    month = {6},
    pages = {A148},
    volume = {566},
    doi = {10.1051/0004-6361/201323097},
    issn = {0004-6361}
}

@article{Kerr2021HeIonizations,
    title = {{He i 10830 {\AA} Dimming during Solar Flares. I. The Crucial Role of Nonthermal Collisional Ionizations}},
    year = {2021},
    journal = {The Astrophysical Journal},
    author = {Kerr, Graham S. and Xu, Yan and Allred, Joel C. and Polito, Vanessa and Sadykov, Viacheslav M. and Huang, Nengyi and Wang, Haimin},
    number = {2},
    month = {5},
    pages = {153},
    volume = {912},
    doi = {10.3847/1538-4357/abf42d},
    issn = {0004-637X}
}

@article{Qiu2012HEATINGFUNCTIONS,
    title = {{HEATING OF FLARE LOOPS WITH OBSERVATIONALLY CONSTRAINED HEATING FUNCTIONS}},
    year = {2012},
    journal = {The Astrophysical Journal},
    author = {Qiu, Jiong and Liu, Wen-Juan and Longcope, Dana W.},
    number = {2},
    month = {6},
    pages = {124},
    volume = {752},
    doi = {10.1088/0004-637X/752/2/124},
    issn = {0004-637X}
}

@inproceedings{Ramaty2000HighFlares,
    title = {{High energy processes in solar flares}},
    year = {2000},
    booktitle = {AIP Conference Proceedings},
    author = {Ramaty, Reuven and Mandzhavidze, Natalie},
    pages = {401--410},
    publisher = {AIP},
    doi = {10.1063/1.1291742},
    issn = {0094243X}
}

@article{Kitahara1990High-resolutionFlare,
    title = {{High-resolution observation and detailed photometry of a great H? two-ribbon flare}},
    year = {1990},
    journal = {Solar Physics},
    author = {Kitahara, T. and Kurokawa, H.},
    number = {2},
    pages = {321--332},
    volume = {125},
    doi = {10.1007/BF00158409},
    issn = {0038-0938}
}

@article{ThoenFaber2025High-resolutionStructures,
    title = {{High-resolution observational analysis of flare ribbon fine structures}},
    year = {2025},
    journal = {Astronomy {\&} Astrophysics},
    author = {Thoen Faber, Jonas and Joshi, Reetika and van der Voort, Luc Rouppe and Wedemeyer, Sven and Fletcher, Lyndsay and Aulanier, Guillaume and N{\'{o}}brega-Siverio, Daniel},
    month = {1},
    pages = {A8},
    volume = {693},
    doi = {10.1051/0004-6361/202452370},
    issn = {0004-6361}
}

@article{DeVore2008HomologousBreakout,
    title = {{Homologous Confined Filament Eruptions via Magnetic Breakout}},
    year = {2008},
    journal = {The Astrophysical Journal},
    author = {DeVore, C. Richard and Antiochos, Spiro K.},
    number = {1},
    month = {6},
    pages = {740--756},
    volume = {680},
    doi = {10.1086/588011},
    issn = {0004-637X}
}

@article{Afanasyev2023HybridRegion,
    title = {{Hybrid Data-driven Magnetofrictional and Magnetohydrodynamic Simulations of an Eruptive Solar Active Region}},
    year = {2023},
    journal = {The Astrophysical Journal},
    author = {Afanasyev, Andrey N. and Fan, Yuhong and Kazachenko, Maria D. and Cheung, Mark C. M.},
    number = {2},
    month = {8},
    pages = {136},
    volume = {952},
    doi = {10.3847/1538-4357/acd7e9},
    issn = {0004-637X}
}

@article{Li2017ImagingFlare,
    title = {{Imaging Observations of Magnetic Reconnection in a Solar Eruptive Flare}},
    year = {2017},
    journal = {The Astrophysical Journal},
    author = {Li, Y. and Sun, X. and Ding, M. D. and Qiu, J. and Priest, E. R.},
    number = {2},
    month = {2},
    pages = {190},
    volume = {835},
    doi = {10.3847/1538-4357/835/2/190},
    issn = {0004-637X}
}

@article{CorchadoAlbelo2024InferringRates,
    title = {{Inferring Fundamental Properties of the Flare Current Sheet Using Flare Ribbons: Oscillations in the Reconnection Flux Rates}},
    year = {2024},
    journal = {The Astrophysical Journal},
    author = {Corchado Albelo, Marcel F. and Kazachenko, Maria D. and Lynch, Benjamin J.},
    number = {1},
    month = {4},
    pages = {16},
    volume = {965},
    url = {https://iopscience.iop.org/article/10.3847/1538-4357/ad25f4},
    doi = {10.3847/1538-4357/ad25f4},
    issn = {0004-637X}
}

@article{Loureiro2007InstabilityChains,
    title = {{Instability of current sheets and formation of plasmoid chains}},
    year = {2007},
    journal = {Physics of Plasmas},
    author = {Loureiro, N. F. and Schekochihin, A. A. and College, King's and Cowley, S. C.},
    number = {10},
    volume = {14},
    doi = {10.1063/1.2783986},
    issn = {1070664X},
    arxivId = {astro-ph/0703631}
}

@article{Kazachenko2022InvitedMagnetism,
    title = {{Invited Review: Short-term Variability with the Observations from the Helioseismic and Magnetic Imager (HMI) Onboard the Solar Dynamics Observatory (SDO): Insights into Flare Magnetism}},
    year = {2022},
    journal = {Solar Physics},
    author = {Kazachenko, Maria D. and Albelo-Corchado, Marcel F. and Tamburri, Cole A. and Welsch, Brian T.},
    number = {5},
    month = {5},
    pages = {59},
    volume = {297},
    url = {https://link.springer.com/10.1007/s11207-022-01987-6},
    doi = {10.1007/s11207-022-01987-6},
    issn = {0038-0938}
}

@article{Wyper2021IsSheet,
    title = {{Is Flare Ribbon Fine Structure Related to Tearing in the Flare Current Sheet?}},
    year = {2021},
    journal = {The Astrophysical Journal},
    author = {Wyper, P. F. and Pontin, D. I.},
    number = {2},
    month = {10},
    pages = {102},
    volume = {920},
    publisher = {American Astronomical Society},
    doi = {10.3847/1538-4357/ac1943},
    issn = {0004-637X},
    arxivId = {2108.10966}
}

@article{Zhao2022LaboratoryRegime,
    title = {{Laboratory observation of plasmoid-dominated magnetic reconnection in hybrid collisional-collisionless regime}},
    year = {2022},
    journal = {Communications Physics},
    author = {Zhao, Zhonghai and An, Honghai and Xie, Yu and Lei, Zhu and Yao, Weipeng and Yuan, Wenqiang and Xiong, Jun and Wang, Chen and Ye, Junjian and Xie, Zhiyong and Fang, Zhiheng and Lei, Anle and Pei, Wenbing and He, Xiantu and Zhou, Weimin and Wang, Wei and Zhu, Shaoping and Qiao, Bin},
    number = {1},
    month = {10},
    pages = {247},
    volume = {5},
    doi = {10.1038/s42005-022-01026-7},
    issn = {2399-3650}
}

@article{Arnold2019Large-scaleModel,
    title = {{Large-scale parallel electric fields and return currents in a global simulation model}},
    year = {2019},
    journal = {Physics of Plasmas},
    author = {Arnold, H. and Drake, J. F. and Swisdak, M. and Dahlin, J.},
    number = {10},
    month = {10},
    volume = {26},
    doi = {10.1063/1.5120373},
    issn = {1070-664X}
}

@article{Collier2024LocalisingFlare,
    title = {{Localising pulsations in the hard X-ray and microwave emission of an X-class flare}},
    year = {2024},
    journal = {Astronomy {\&} Astrophysics},
    author = {Collier, Hannah and Hayes, Laura A. and Yu, Sijie and Battaglia, Andrea F. and Ashfield, William and Polito, Vanessa and Harra, Louise K. and Krucker, Säm},
    month = {4},
    pages = {A215},
    volume = {684},
    doi = {10.1051/0004-6361/202348652},
    issn = {0004-6361}
}

@article{Priest1992MagneticPoints,
    title = {{Magnetic flipping: Reconnection in three dimensions without null points}},
    year = {1992},
    journal = {Journal of Geophysical Research: Space Physics},
    author = {Priest, E. R. and Forbes, T. G.},
    number = {A2},
    month = {2},
    pages = {1521--1531},
    volume = {97},
    doi = {10.1029/91JA02435},
    issn = {0148-0227}
}

@article{Qiu2004MagneticEvents,
    title = {{Magnetic Reconnection and Mass Acceleration in Flare–Coronal Mass Ejection Events}},
    year = {2004},
    journal = {The Astrophysical Journal},
    author = {Qiu, Jiong and Wang, Haimin and Cheng, C. Z. and Gary, Dale E.},
    number = {2},
    month = {4},
    pages = {900--905},
    volume = {604},
    doi = {10.1086/382122},
    issn = {0004-637X}
}

@article{Drake2025MagneticWind,
    title = {{Magnetic Reconnection in Solar Flares and the Near-Sun Solar Wind}},
    year = {2025},
    journal = {Space Science Reviews},
    author = {Drake, J. F. and Antiochos, S. K. and Bale, S. D. and Chen, Bin and Cohen, C. M. S. and Dahlin, J. T. and Glesener, Lindsay and Guo, F. and Hoshino, M. and Imada, Shinsuke and Oka, M. and Phan, T. D. and Reeves, Katherine K. and Swisdak, M.},
    number = {2},
    month = {3},
    pages = {27},
    volume = {221},
    doi = {10.1007/s11214-025-01153-x},
    issn = {0038-6308}
}

@article{Kopp1976MagneticPhenomenon,
    title = {{Magnetic reconnection in the corona and the loop prominence phenomenon}},
    year = {1976},
    journal = {Solar Physics},
    author = {Kopp, R.A. and Pneuman, G.W.},
    number = {1},
    volume = {50},
    doi = {10.1007/BF00206193},
    issn = {0038-0938}
}

@article{Cannon2023Magnetic22,
    title = {{Magnetic Reconnection Rate in the M6.5 Solar Flare on 2015 June 22}},
    year = {2023},
    journal = {The Astrophysical Journal},
    author = {Cannon, Bryce and Jing, Ju and Li, Qin and Liu, Nian and Lee, Jeongwoo and Cao, Wenda and Wang, Haimin},
    number = {2},
    month = {6},
    pages = {144},
    volume = {950},
    doi = {10.3847/1538-4357/accf9f},
    issn = {0004-637X}
}

@article{Guidoni2016MAGNETIC-ISLANDFLARES,
    title = {{MAGNETIC-ISLAND CONTRACTION AND PARTICLE ACCELERATION IN SIMULATED ERUPTIVE SOLAR FLARES}},
    year = {2016},
    journal = {The Astrophysical Journal},
    author = {Guidoni, S. E. and DeVore, C. R. and Karpen, J. T. and Lynch, B. J.},
    number = {1},
    month = {3},
    pages = {60},
    volume = {820},
    publisher = {American Astronomical Society},
    doi = {10.3847/0004-637x/820/1/60},
    issn = {15384357}
}

@article{Bobra2021Mbobra/SHARPs:2021-07-23,
    title = {{mbobra/SHARPs: SHARPs 0.1.0 (2021-07-23)}},
    year = {2021},
    author = {Bobra, Monica G. and Xudong, Sun and Turmon, Michael J.},
    month = {7},
    url = {https://doi.org/10.5281/zenodo.5131292#.Yy4sNfNuIIw.mendeley},
    doi = {10.5281/ZENODO.5131292}
}

@article{Grassberger1983MeasuringAttractors,
    title = {{Measuring the strangeness of strange attractors}},
    year = {1983},
    journal = {Physica D: Nonlinear Phenomena},
    author = {Grassberger, Peter and Procaccia, Itamar},
    number = {1-2},
    month = {10},
    pages = {189--208},
    volume = {9},
    doi = {10.1016/0167-2789(83)90298-1},
    issn = {01672789}
}

@article{Sturrock1966ModelFlares,
    title = {{Model of the High-Energy Phase of Solar Flares}},
    year = {1966},
    journal = {Nature},
    author = {Sturrock, P. A.},
    number = {5050},
    month = {8},
    pages = {695--697},
    volume = {211},
    url = {https://www.nature.com/articles/211695a0},
    doi = {10.1038/211695a0},
    issn = {0028-0836}
}

@article{Parker2017ModelingShear,
    title = {{Modeling a Propagating Sawtooth Flare Ribbon Structure as a Tearing Mode in the Presence of Velocity Shear}},
    year = {2017},
    journal = {The Astrophysical Journal},
    author = {Parker, Jacob and Longcope, Dana},
    number = {1},
    month = {9},
    pages = {30},
    volume = {847},
    doi = {10.3847/1538-4357/aa8908},
    issn = {1538-4357}
}

@article{Longcope2007Modeling2004,
    title = {{Modeling and Measuring the Flux Reconnected and Ejected by the Two-Ribbon Flare/CME Event on 7 November 2004}},
    year = {2007},
    journal = {Solar Physics},
    author = {Longcope, Dana and Beveridge, Colin and Qiu, Jiong and Ravindra, B. and Barnes, Graham and Dasso, Sergio},
    number = {1-2},
    month = {10},
    pages = {45--73},
    volume = {244},
    doi = {10.1007/s11207-007-0330-7},
    issn = {0038-0938}
}

@article{Yadav2025Multi-lineViSP/DKIST,
    title = {{Multi-line Spectropolarimetric Observation of Flare Ribbon Fine Structures with ViSP/DKIST}},
    year = {2025},
    journal = {The Astrophysical Journal},
    author = {Yadav, Rahul and Kazachenko, Maria D. and Cauzzi, Gianna and Tamburri, Cole and Corchado, Marcel and French, Ryan},
    doi = {10.3847/1538-4357/adf4c1},
    arxivId = {2507.20070}
}

@inproceedings{Chamberlin2009NextSeries,
    title = {{Next generation x-ray sensor (XRS) for the NOAA GOES-R satellite series}},
    year = {2009},
    booktitle = {Solar Physics and Space Weather Instrumentation III},
    author = {Chamberlin, Phillip C. and Woods, Thomas N. and Eparvier, Francis G. and Jones, Andrew R.},
    editor = {Fineschi, Silvano and Fennelly, Judy A.},
    month = {8},
    pages = {743802},
    volume = {7438},
    publisher = {SPIE},
    url = {http://proceedings.spiedigitallibrary.org/proceeding.aspx?doi=10.1117/12.826807},
    doi = {10.1117/12.826807}
}

@article{Shen2023Non-thermalEruptions,
    title = {{Non-thermal broadening of IRIS Fe XXI line caused by turbulent plasma flows in the magnetic reconnection region during solar eruptions}},
    year = {2023},
    journal = {Frontiers in Astronomy and Space Sciences},
    author = {Shen, Chengcai and Polito, Vanessa and Reeves, Katharine K. and Chen, Bin and Yu, Sijie and Xie, Xiaoyan},
    month = {2},
    volume = {10},
    doi = {10.3389/fspas.2023.1096133},
    issn = {2296-987X}
}

@article{Ashfield2024NonthermalAcceleration,
    title = {{Nonthermal Observations of a Flare Loop-top Using IRIS Fe xxi: Implications for Turbulence and Electron Acceleration}},
    year = {2024},
    journal = {The Astrophysical Journal},
    author = {Ashfield, William and Polito, Vanessa and Yu, Sijie and Collier, Hannah and Hayes, Laura A.},
    number = {2},
    month = {10},
    pages = {96},
    volume = {973},
    doi = {10.3847/1538-4357/ad64ca},
    issn = {0004-637X}
}

@article{Magara1996NumericalFlares,
    title = {{Numerical Simulation of Magnetic Reconnection in Eruptive Flares}},
    year = {1996},
    journal = {The Astrophysical Journal},
    author = {Magara, Tetsuya and Mineshige, Shin and Yokoyama, Takaaki and Shibata, Kazunari},
    month = {8},
    pages = {1054},
    volume = {466},
    doi = {10.1086/177575},
    issn = {0004-637X}
}

@article{Forbes1984NumericalRegion,
    title = {{Numerical simulation of reconnection in an emerging magnetic flux region}},
    year = {1984},
    journal = {Solar Physics},
    author = {Forbes, T. G. and Priest, E. R.},
    number = {2},
    month = {9},
    pages = {315--340},
    volume = {94},
    doi = {10.1007/BF00151321},
    issn = {0038-0938}
}

@article{Shibata2023NumericalFlare,
    title = {{Numerical Study on Excitation of Turbulence and Oscillation in Above-the-loop-top Region of a Solar Flare}},
    year = {2023},
    journal = {The Astrophysical Journal},
    author = {Shibata, Kengo and Takasao, Shinsuke and Reeves, Katharine K.},
    number = {2},
    month = {2},
    pages = {106},
    volume = {943},
    doi = {10.3847/1538-4357/acaa9c},
    issn = {0004-637X}
}

@article{Lu2022ObservationalFlare,
    title = {{Observational Signatures of Tearing Instability in the Current Sheet of a Solar Flare}},
    year = {2022},
    journal = {The Astrophysical Journal Letters},
    author = {Lu, Lei and Feng, Li and Warmuth, Alexander and Veronig, Astrid M. and Huang, Jing and Liu, Siming and Gan, Weiqun and Ning, Zongjun and Ying, Beili and Gao, Guannan},
    number = {1},
    month = {1},
    pages = {L7},
    volume = {924},
    doi = {10.3847/2041-8213/ac42c6},
    issn = {2041-8205}
}

@article{Schmidt2025ObservationsOptics,
    title = {{Observations of fine coronal structures with high-order solar adaptive optics}},
    year = {2025},
    journal = {Nature Astronomy},
    author = {Schmidt, Dirk and Schad, Thomas A. and Yurchyshyn, Vasyl and Gorceix, Nicolas and Rimmele, Thomas R. and Goode, Philip R.},
    month = {5},
    doi = {10.1038/s41550-025-02564-0},
    issn = {2397-3366}
}

@article{Sun2013OnNote,
    title = {{On the Coordinate System of Space-Weather HMI Active Region Patches (SHARPs): A Technical Note}},
    year = {2013},
    author = {Sun, Xudong},
    month = {9},
    url = {http://arxiv.org/abs/1309.2392},
    arxivId = {1309.2392}
}

@article{Cho2024OnIRIS,
    title = {{On the Nature of Nonthermal Broadening of Spectral Lines Observed by IRIS}},
    year = {2024},
    journal = {The Astrophysical Journal},
    author = {Cho, Kyuhyoun and De Pontieu, Bart and Testa, Paola},
    number = {1},
    month = {11},
    pages = {33},
    volume = {975},
    doi = {10.3847/1538-4357/ad7586},
    issn = {0004-637X}
}

@article{Hesse2005OnCorona,
    title = {{On the Relation between Reconnected Magnetic Flux and Parallel Electric Fields in the Solar Corona}},
    year = {2005},
    journal = {The Astrophysical Journal},
    author = {Hesse, Michael and Forbes, Terry G. and Birn, Joachim},
    number = {2},
    month = {10},
    pages = {1227--1238},
    volume = {631},
    doi = {10.1086/432677},
    issn = {0004-637X}
}

@article{Nakamura2025OutstandingReconnection,
    title = {{Outstanding Questions and Future Research on Magnetic Reconnection}},
    year = {2025},
    journal = {Space Science Reviews},
    author = {Nakamura, R. and Burch, J. L. and Birn, J. and Chen, L.-J. and Graham, D. B. and Guo, F. and Hwang, K.-J. and Ji, H. and Khotyaintsev, Y. V. and Liu, Y.-H. and Oka, M. and Payne, D. and Sitnov, M. I. and Swisdak, M. and Zenitani, S. and Drake, J. F. and Fuselier, S. A. and Genestreti, K. J. and Gershman, D. J. and Hasegawa, H. and Hoshino, M. and Norgren, C. and Shay, M. A. and Shuster, J. R. and Stawarz, J. E.},
    number = {1},
    month = {2},
    pages = {17},
    volume = {221},
    doi = {10.1007/s11214-025-01143-z},
    issn = {0038-6308}
}

@article{Keppens2012ParallelMagnetohydrodynamics,
    title = {{Parallel, grid-adaptive approaches for relativistic hydro and magnetohydrodynamics}},
    year = {2012},
    journal = {Journal of Computational Physics},
    author = {Keppens, R. and Meliani, Z. and van Marle, A.J. and Delmont, P. and Vlasis, A. and van der Holst, B.},
    number = {3},
    month = {2},
    pages = {718--744},
    volume = {231},
    doi = {10.1016/j.jcp.2011.01.020},
    issn = {00219991}
}

@article{Chen2015ParticleShock,
    title = {{Particle acceleration by a solar flare termination shock}},
    year = {2015},
    journal = {Science},
    author = {Chen, Bin and Bastian, Timothy S. and Shen, Chengcai and Gary, Dale E. and Krucker, Säm and Glesener, Lindsay},
    number = {6265},
    month = {12},
    pages = {1238--1242},
    volume = {350},
    doi = {10.1126/science.aac8467},
    issn = {0036-8075}
}

@article{Bacchini2024ParticleLoops,
    title = {{Particle trapping and acceleration in turbulent post-flare coronal loops}},
    year = {2024},
    journal = {Monthly Notices of the Royal Astronomical Society},
    author = {Bacchini, Fabio and Ruan, Wenzhi and Keppens, Rony},
    number = {3},
    month = {3},
    pages = {2399--2412},
    volume = {529},
    doi = {10.1093/mnras/stae723},
    issn = {0035-8711}
}

@article{Comisso2017PlasmoidSheets,
    title = {{Plasmoid Instability in Forming Current Sheets}},
    year = {2017},
    journal = {The Astrophysical Journal},
    author = {Comisso, L. and Lingam, M. and Huang, Y.-M. and Bhattacharjee, A.},
    number = {2},
    month = {12},
    pages = {142},
    volume = {850},
    doi = {10.3847/1538-4357/aa9789},
    issn = {0004-637X}
}

@article{Fielding2023PlasmoidMedium,
    title = {{Plasmoid Instability in the Multiphase Interstellar Medium}},
    year = {2023},
    journal = {The Astrophysical Journal Letters},
    author = {Fielding, Drummond B. and Ripperda, Bart and Philippov, Alexander A.},
    number = {1},
    month = {5},
    pages = {L5},
    volume = {949},
    doi = {10.3847/2041-8213/accf1f},
    issn = {2041-8205}
}

@article{Shibata2001Plasmoid-induced-reconnectionReconnection,
    title = {{Plasmoid-induced-reconnection and fractal reconnection}},
    year = {2001},
    journal = {Earth, Planets and Space},
    author = {Shibata, Kazunari and Tanuma, Syuniti},
    number = {6},
    month = {6},
    pages = {473--482},
    volume = {53},
    url = {http://earth-planets-space.springeropen.com/articles/10.1186/BF03353258},
    doi = {10.1186/BF03353258},
    issn = {1880-5981}
}

@article{French2021ProbingDynamics,
    title = {{Probing Current Sheet Instabilities from Flare Ribbon Dynamics}},
    year = {2021},
    journal = {The Astrophysical Journal},
    author = {French, Ryan J. and Matthews, Sarah A. and Jonathan Rae, I. and Smith, Andrew W.},
    number = {2},
    month = {12},
    pages = {117},
    volume = {922},
    doi = {10.3847/1538-4357/ac256f},
    issn = {0004-637X}
}

@article{Wyper2015QuantifyingLayers,
    title = {{Quantifying three dimensional reconnection in fragmented current layers}},
    year = {2015},
    journal = {Physics of Plasmas},
    author = {Wyper, P. F. and Hesse, M.},
    number = {4},
    month = {4},
    volume = {22},
    doi = {10.1063/1.4918335},
    issn = {1070-664X}
}

@article{Clarke2021Quasi-periodicFlare,
    title = {{Quasi-periodic Particle Acceleration in a Solar Flare}},
    year = {2021},
    journal = {The Astrophysical Journal},
    author = {Clarke, Brendan P. and Hayes, Laura A. and Gallagher, Peter T. and Maloney, Shane A. and Carley, Eoin P.},
    number = {2},
    month = {4},
    pages = {123},
    volume = {910},
    doi = {10.3847/1538-4357/abe463},
    issn = {0004-637X}
}

@article{Lorincik2022RapidCadence,
    title = {{Rapid variations of Si IV spectra in a flare observed by interface region imaging spectrograph at a sub-second cadence}},
    year = {2022},
    journal = {Frontiers in Astronomy and Space Sciences},
    author = {L{\"{o}}rin{\v{c}}{\'{i}}k, Juraj and Polito, Vanessa and De Pontieu, Bart and Yu, Sijie and Freij, Nabil},
    month = {11},
    volume = {9},
    doi = {10.3389/fspas.2022.1040945},
    issn = {2296-987X}
}

@article{Lynch2016RECONNECTIONERUPTIONS,
    title = {{RECONNECTION PROPERTIES OF LARGE-SCALE CURRENT SHEETS DURING CORONAL MASS EJECTION ERUPTIONS}},
    year = {2016},
    journal = {The Astrophysical Journal},
    author = {Lynch, B. J. and Edmondson, J. K. and Kazachenko, M. D. and Guidoni, S. E.},
    number = {1},
    month = {7},
    pages = {43},
    volume = {826},
    doi = {10.3847/0004-637X/826/1/43},
    issn = {1538-4357}
}

@article{Daughton2011RolePlasmas,
    title = {{Role of electron physics in the development of turbulent magnetic reconnection in collisionless plasmas}},
    year = {2011},
    journal = {Nature Physics},
    author = {Daughton, W. and Roytershteyn, V. and Karimabadi, H. and Yin, L. and Albright, B. J. and Bergen, B. and Bowers, K. J.},
    number = {7},
    month = {7},
    pages = {539--542},
    volume = {7},
    doi = {10.1038/nphys1965},
    issn = {1745-2473}
}

@article{Virtanen2020SciPyPython,
    title = {{SciPy 1.0: fundamental algorithms for scientific computing in Python}},
    year = {2020},
    journal = {Nature Methods},
    author = {Virtanen, Pauli and Gommers, Ralf and Oliphant, Travis E. and Haberland, Matt and Reddy, Tyler and Cournapeau, David and Burovski, Evgeni and Peterson, Pearu and Weckesser, Warren and Bright, Jonathan and van der Walt, Stéfan J. and Brett, Matthew and Wilson, Joshua and Millman, K. Jarrod and Mayorov, Nikolay and Nelson, Andrew R. J. and Jones, Eric and Kern, Robert and Larson, Eric and Carey, C J and Polat, İlhan and Feng, Yu and Moore, Eric W. and VanderPlas, Jake and Laxalde, Denis and Perktold, Josef and Cimrman, Robert and Henriksen, Ian and Quintero, E. A. and Harris, Charles R. and Archibald, Anne M. and Ribeiro, Antônio H. and Pedregosa, Fabian and van Mulbregt, Paul and Vijaykumar, Aditya and Bardelli, Alessandro Pietro and Rothberg, Alex and Hilboll, Andreas and Kloeckner, Andreas and Scopatz, Anthony and Lee, Antony and Rokem, Ariel and Woods, C. Nathan and Fulton, Chad and Masson, Charles and H{\"{a}}ggstr{\"{o}}m, Christian and Fitzgerald, Clark and Nicholson, David A. and Hagen, David R. and Pasechnik, Dmitrii V. and Olivetti, Emanuele and Martin, Eric and Wieser, Eric and Silva, Fabrice and Lenders, Felix and Wilhelm, Florian and Young, G. and Price, Gavin A. and Ingold, Gert-Ludwig and Allen, Gregory E. and Lee, Gregory R. and Audren, Hervé and Probst, Irvin and Dietrich, Jörg P. and Silterra, Jacob and Webber, James T and Slavi{\v{c}}, Janko and Nothman, Joel and Buchner, Johannes and Kulick, Johannes and Sch{\"{o}}nberger, Johannes L. and de Miranda Cardoso, José Vinícius and Reimer, Joscha and Harrington, Joseph and Rodr{\'{i}}guez, Juan Luis Cano and Nunez-Iglesias, Juan and Kuczynski, Justin and Tritz, Kevin and Thoma, Martin and Newville, Matthew and K{\"{u}}mmerer, Matthias and Bolingbroke, Maximilian and Tartre, Michael and Pak, Mikhail and Smith, Nathaniel J. and Nowaczyk, Nikolai and Shebanov, Nikolay and Pavlyk, Oleksandr and Brodtkorb, Per A. and Lee, Perry and McGibbon, Robert T. and Feldbauer, Roman and Lewis, Sam and Tygier, Sam and Sievert, Scott and Vigna, Sebastiano and Peterson, Stefan and More, Surhud and Pudlik, Tadeusz and Oshima, Takuya and Pingel, Thomas J. and Robitaille, Thomas P. and Spura, Thomas and Jones, Thouis R. and Cera, Tim and Leslie, Tim and Zito, Tiziano and Krauss, Tom and Upadhyay, Utkarsh and Halchenko, Yaroslav O. and V{\'{a}}zquez-Baeza, Yoshiki},
    number = {3},
    month = {3},
    pages = {261--272},
    volume = {17},
    doi = {10.1038/s41592-019-0686-2},
    issn = {1548-7091}
}

@article{Kerr2019SISimulations,
    title = {{SI iv Resonance Line Emission during Solar Flares: Non-LTE, Nonequilibrium, Radiation Transfer Simulations}},
    year = {2019},
    journal = {The Astrophysical Journal},
    author = {Kerr, Graham S. and Carlsson, Mats and Allred, Joel C. and Young, Peter R. and Daw, Adrian N.},
    number = {1},
    month = {1},
    pages = {23},
    volume = {871},
    doi = {10.3847/1538-4357/aaf46e},
    issn = {0004-637X}
}

@article{Savcheva2012SIGMOIDALMODEL,
    title = {{SIGMOIDAL ACTIVE REGION ON THE SUN: COMPARISON OF A MAGNETOHYDRODYNAMICAL SIMULATION AND A NONLINEAR FORCE-FREE FIELD MODEL}},
    year = {2012},
    journal = {The Astrophysical Journal},
    author = {Savcheva, A. and Pariat, E. and van Ballegooijen, A. and Aulanier, G. and DeLuca, E.},
    number = {1},
    month = {5},
    pages = {15},
    volume = {750},
    doi = {10.1088/0004-637X/750/1/15},
    issn = {0004-637X}
}

@article{Takasao2012SIMULTANEOUSFLARE,
    title = {{SIMULTANEOUS OBSERVATION OF RECONNECTION INFLOW AND OUTFLOW ASSOCIATED WITH THE 2010 AUGUST 18 SOLAR FLARE}},
    year = {2012},
    journal = {The Astrophysical Journal},
    author = {Takasao, Shinsuke and Asai, Ayumi and Isobe, Hiroaki and Shibata, Kazunari},
    number = {1},
    month = {1},
    pages = {L6},
    volume = {745},
    doi = {10.1088/2041-8205/745/1/L6},
    issn = {2041-8205}
}

@article{Kliem2000SolarReconnection,
    title = {{Solar flare radio pulsations as a signature of dynamic magnetic reconnection}},
    year = {2000},
    journal = {Astron. Astrophys},
    author = {Kliem, B and Karlick{\'{y}}, M and Benz, A O},
    pages = {715--728},
    volume = {360},
    url = {https://ui.adsabs.harvard.edu/abs/2000A%26A...360..715K/abstract}
}

@article{Polito2023SolarSpectroscopy,
    title = {{Solar Flare Ribbon Fronts. I. Constraining Flare Energy Deposition with IRIS Spectroscopy}},
    year = {2023},
    journal = {The Astrophysical Journal},
    author = {Polito, Vanessa and Kerr, Graham S. and Xu, Yan and Sadykov, Viacheslav M. and Lorincik, Juraj},
    number = {1},
    month = {2},
    pages = {104},
    volume = {944},
    doi = {10.3847/1538-4357/acaf7c},
    issn = {0004-637X}
}

@article{Kerr2024SolarFootpoints,
    title = {{Solar Flare Ribbon Fronts. II. Evolution of Heating Rates in Individual Flare Footpoints}},
    year = {2024},
    journal = {The Astrophysical Journal},
    author = {Kerr, Graham S. and Polito, Vanessa and Xu, Yan and Allred, Joel C.},
    number = {1},
    month = {7},
    pages = {21},
    volume = {970},
    doi = {10.3847/1538-4357/ad47e1},
    issn = {0004-637X}
}

@article{Shibata2011SolarProcesses,
    title = {{Solar Flares: Magnetohydrodynamic Processes}},
    year = {2011},
    journal = {Living Reviews in Solar Physics},
    author = {Shibata, Kazunari and Magara, Tetsuya},
    volume = {8},
    url = {http://link.springer.com/10.12942/lrsp-2011-6},
    doi = {10.12942/lrsp-2011-6},
    issn = {1614-4961}
}

@article{Karlicky2014SolarProcesses,
    title = {{Solar flares: radio and X-ray signatures of magnetic reconnection processes}},
    year = {2014},
    journal = {Research in Astronomy and Astrophysics},
    author = {Karlick{\'{y}}, Marian},
    number = {7},
    month = {7},
    pages = {753--772},
    volume = {14},
    doi = {10.1088/1674-4527/14/7/002},
    issn = {1674-4527}
}

@article{Purkhart2025SpatiotemporalAcceleration,
    title = {{Spatiotemporal evolution of UV pulsations and their connection to 3D magnetic reconnection and particle acceleration}},
    year = {2025},
    journal = {Astronomy {\&} Astrophysics},
    author = {Purkhart, Stefan and Collier, Hannah and Hayes, Laura A. and Veronig, Astrid M. and Janvier, Miho and Krucker, Säm},
    month = {6},
    pages = {A318},
    volume = {698},
    doi = {10.1051/0004-6361/202554475},
    issn = {0004-6361}
}

@article{Guidoni2022SpectralIslands,
    title = {{Spectral Power-law Formation by Sequential Particle Acceleration in Multiple Flare Magnetic Islands}},
    year = {2022},
    journal = {The Astrophysical Journal},
    author = {Guidoni, S. E. and Karpen, J. T. and DeVore, C. R.},
    number = {2},
    month = {2},
    pages = {191},
    volume = {925},
    doi = {10.3847/1538-4357/ac39a5},
    issn = {0004-637X}
}

@article{Pietrow2024SpectralRibbons,
    title = {{Spectral variations within solar flare ribbons}},
    year = {2024},
    journal = {Astronomy {\&} Astrophysics},
    author = {Pietrow, A. G. M. and Druett, M. K. and Singh, V.},
    month = {5},
    pages = {A137},
    volume = {685},
    doi = {10.1051/0004-6361/202348839},
    issn = {0004-6361}
}

@article{Brannon2015SPECTROSCOPICWAVES,
    title = {{SPECTROSCOPIC OBSERVATIONS OF AN EVOLVING FLARE RIBBON SUBSTRUCTURE SUGGESTING ORIGIN IN CURRENT SHEET WAVES}},
    year = {2015},
    journal = {The Astrophysical Journal},
    author = {Brannon, S. R. and Longcope, D. W. and Qiu, J.},
    number = {1},
    month = {8},
    pages = {4},
    volume = {810},
    publisher = {Institute of Physics Publishing},
    url = {https://iopscience.iop.org/article/10.1088/0004-637X/810/1/4},
    doi = {10.1088/0004-637X/810/1/4},
    issn = {1538-4357}
}

@article{Barta2011SPONTANEOUSOBSERVATIONS,
    title = {{SPONTANEOUS CURRENT-LAYER FRAGMENTATION AND CASCADING RECONNECTION IN SOLAR FLARES. II. RELATION TO OBSERVATIONS}},
    year = {2011},
    journal = {The Astrophysical Journal},
    author = {B{\'{a}}rta, Miroslav and B{\"{u}}chner, Jörg and Karlick{\'{y}}, Marian and Kotr{\v{c}}, Pavel},
    number = {1},
    month = {3},
    pages = {47},
    volume = {730},
    doi = {10.1088/0004-637X/730/1/47},
    issn = {0004-637X}
}

@article{Mason2022StatisticalBoundary,
    title = {{Statistical Evidence for Small-scale Interchange Reconnection at a Coronal Hole Boundary}},
    year = {2022},
    journal = {The Astrophysical Journal Letters},
    author = {Mason, Emily I. and Uritsky, Vadim M.},
    number = {1},
    month = {9},
    pages = {L19},
    volume = {937},
    doi = {10.3847/2041-8213/ac9124},
    issn = {2041-8205}
}

@article{Gopalswamy2018Sun-to-earthObservations,
    title = {{Sun-to-earth propagation of the 2015 June 21 coronal mass ejection revealed by optical, EUV, and radio observations}},
    year = {2018},
    journal = {Journal of Atmospheric and Solar-Terrestrial Physics},
    author = {Gopalswamy, N. and Makela, P. and Akiyama, S. and Yashiro, S. and Xie, H. and Thakur, N.},
    month = {11},
    pages = {225--238},
    volume = {179},
    doi = {10.1016/j.jastp.2018.07.013},
    issn = {13646826}
}

@article{Bouligand1929SurPlan,
    title = {{Sur la notion d’ordre de mesure d’un ensemble plan}},
    year = {1929},
    journal = {Bull. Sci. Math},
    author = {Bouligand, G},
    pages = {185--192},
    volume = {2}
}

@article{Alexander2006TemporalFlares,
    title = {{Temporal and Spatial Relationships between Ultraviolet and Hard X‐Ray Emission in Solar Flares}},
    year = {2006},
    journal = {The Astrophysical Journal},
    author = {Alexander, David and Coyner, Aaron J.},
    number = {1},
    month = {3},
    pages = {505--515},
    volume = {640},
    doi = {10.1086/500076},
    issn = {0004-637X}
}

@article{Graham2015TEMPORALFLARE,
    title = {{TEMPORAL EVOLUTION OF MULTIPLE EVAPORATING RIBBON SOURCES IN A SOLAR FLARE}},
    year = {2015},
    journal = {The Astrophysical Journal},
    author = {Graham, D. R. and Cauzzi, G.},
    number = {2},
    month = {7},
    pages = {L22},
    volume = {807},
    doi = {10.1088/2041-8205/807/2/L22},
    issn = {2041-8213}
}

@article{Lemen2012TheSDO,
    title = {{The Atmospheric Imaging Assembly (AIA) on the Solar Dynamics Observatory (SDO)}},
    year = {2012},
    journal = {Solar Physics},
    author = {Lemen, James R. and Title, Alan M. and Akin, David J. and Boerner, Paul F. and Chou, Catherine and Drake, Jerry F. and Duncan, Dexter W. and Edwards, Christopher G. and Friedlaender, Frank M. and Heyman, Gary F. and Hurlburt, Neal E. and Katz, Noah L. and Kushner, Gary D. and Levay, Michael and Lindgren, Russell W. and Mathur, Dnyanesh P. and McFeaters, Edward L. and Mitchell, Sarah and Rehse, Roger A. and Schrijver, Carolus J. and Springer, Larry A. and Stern, Robert A. and Tarbell, Theodore D. and Wuelser, Jean-Pierre and Wolfson, C. Jacob and Yanari, Carl and Bookbinder, Jay A. and Cheimets, Peter N. and Caldwell, David and Deluca, Edward E. and Gates, Richard and Golub, Leon and Park, Sang and Podgorski, William A. and Bush, Rock I. and Scherrer, Philip H. and Gummin, Mark A. and Smith, Peter and Auker, Gary and Jerram, Paul and Pool, Peter and Soufli, Regina and Windt, David L. and Beardsley, Sarah and Clapp, Matthew and Lang, James and Waltham, Nicholas},
    number = {1-2},
    month = {1},
    pages = {17--40},
    volume = {275},
    url = {https://link.springer.com/10.1007/s11207-011-9776-8},
    doi = {10.1007/s11207-011-9776-8},
    issn = {0038-0938}
}

@article{Kowalski2017The29,
    title = {{The Atmospheric Response to High Nonthermal Electron Beam Fluxes in Solar Flares. I. Modeling the Brightest NUV Footpoints in the X1 Solar Flare of 2014 March 29}},
    year = {2017},
    journal = {The Astrophysical Journal},
    author = {Kowalski, Adam F. and Allred, Joel C. and Daw, Adrian and Cauzzi, Gianna and Carlsson, Mats},
    number = {1},
    month = {2},
    pages = {12},
    volume = {836},
    doi = {10.3847/1538-4357/836/1/12},
    issn = {0004-637X}
}

@article{Woger2021TheVBI,
    title = {{The Daniel K. Inouye Solar Telescope (DKIST)/Visible Broadband Imager (VBI)}},
    year = {2021},
    journal = {Solar Physics},
    author = {W{\"{o}}ger, Friedrich and Rimmele, Thomas and Ferayorni, Andrew and Beard, Andrew and Gregory, Brian S. and Sekulic, Predrag and Hegwer, Steven L.},
    number = {10},
    month = {10},
    pages = {145},
    volume = {296},
    doi = {10.1007/s11207-021-01881-7},
    issn = {0038-0938}
}

@article{Rimmele2020TheOverview,
    title = {{The Daniel K. Inouye Solar Telescope – Observatory Overview}},
    year = {2020},
    journal = {Solar Physics},
    author = {Rimmele, Thomas R. and Warner, Mark and Keil, Stephen L. and Goode, Philip R. and Kn{\"{o}}lker, Michael and Kuhn, Jeffrey R. and Rosner, Robert R. and McMullin, Joseph P. and Casini, Roberto and Lin, Haosheng and W{\"{o}}ger, Friedrich and von der L{\"{u}}he, Oskar and Tritschler, Alexandra and Davey, Alisdair and de Wijn, Alfred and Elmore, David F. and Fehlmann, André and Harrington, David M. and Jaeggli, Sarah A. and Rast, Mark P. and Schad, Thomas A. and Schmidt, Wolfgang and Mathioudakis, Mihalis and Mickey, Donald L. and Anan, Tetsu and Beck, Christian and Marshall, Heather K. and Jeffers, Paul F. and Oschmann, Jacobus M. and Beard, Andrew and Berst, David C. and Cowan, Bruce A. and Craig, Simon C. and Cross, Eric and Cummings, Bryan K. and Donnelly, Colleen and de Vanssay, Jean-Benoit and Eigenbrot, Arthur D. and Ferayorni, Andrew and Foster, Christopher and Galapon, Chriselle Ann and Gedrites, Christopher and Gonzales, Kerry and Goodrich, Bret D. and Gregory, Brian S. and Guzman, Stephanie S. and Guzzo, Stephen and Hegwer, Steve and Hubbard, Robert P. and Hubbard, John R. and Johansson, Erik M. and Johnson, Luke C. and Liang, Chen and Liang, Mary and McQuillen, Isaac and Mayer, Christopher and Newman, Karl and Onodera, Brialyn and Phelps, LeEllen and Puentes, Myles M. and Richards, Christopher and Rimmele, Lukas M. and Sekulic, Predrag and Shimko, Stephan R. and Simison, Brett E. and Smith, Brett and Starman, Erik and Sueoka, Stacey R. and Summers, Richard T. and Szabo, Aimee and Szabo, Louis and Wampler, Stephen B. and Williams, Timothy R. and White, Charles},
    number = {12},
    month = {12},
    pages = {172},
    volume = {295},
    url = {http://link.springer.com/10.1007/s11207-020-01736-7},
    doi = {10.1007/s11207-020-01736-7},
    issn = {0038-0938}
}

@article{Brown1971TheBursts,
    title = {{The deduction of energy spectra of non-thermal electrons in flares from the observed dynamic spectra of hard X-ray bursts}},
    year = {1971},
    journal = {Solar Physics},
    author = {Brown, John C.},
    number = {3},
    month = {7},
    pages = {489--502},
    volume = {18},
    doi = {10.1007/BF00149070},
    issn = {0038-0938}
}

@article{Jeffrey2018TheFlare,
    title = {{The development of lower-atmosphere turbulence early in a solar flare}},
    year = {2018},
    journal = {Science Advances},
    author = {Jeffrey, N. L. S. and Fletcher, L. and Labrosse, N. and Simões, P. J. A.},
    number = {12},
    month = {12},
    volume = {4},
    doi = {10.1126/sciadv.aav2794},
    issn = {2375-2548}
}

@article{Meegan2009TheMonitor,
    title = {{The fermi gamma-ray burst monitor}},
    year = {2009},
    journal = {Astrophysical Journal},
    author = {Meegan, Charles and Lichti, Giselher and Bhat, P. N. and Bissaldi, Elisabetta and Briggs, Michael S. and Connaughton, Valerie and Diehl, Roland and Fishman, Gerald and Greiner, Jochen and Hoover, Andrew S. and Van Der Horst, Alexander J. and Von Kienlin, Andreas and Kippen, R. Marc and Kouveliotou, Chryssa and McBreen, Sheila and Paciesas, W. S. and Preece, Robert and Steinle, Helmut and Wallace, Mark S. and Wilson, Robert B. and Wilson-Hodge, Colleen},
    number = {1},
    pages = {791--804},
    volume = {702},
    publisher = {Institute of Physics Publishing},
    doi = {10.1088/0004-637X/702/1/791},
    issn = {15384357}
}

@incollection{Somov2006ThePlasma,
    title = {{The Generalized Ohm’s Law in Plasma}},
    year = {2006},
    booktitle = {Plasma Astrophysics, Part II},
    author = {Somov, Boris},
    chapter = {11},
    pages = {193--204},
    volume = {341},
    publisher = {Springer},
    url = {http://link.springer.com/10.1007/978-0-387-68894-7_12},
    address = {New York},
    doi = {10.1007/978-0-387-68894-7{\_}12}
}

@article{Hoeksema2014ThePerformance,
    title = {{The Helioseismic and Magnetic Imager (HMI) Vector Magnetic Field Pipeline: Overview and Performance}},
    year = {2014},
    journal = {Solar Physics},
    author = {Hoeksema, J. Todd and Liu, Yang and Hayashi, Keiji and Sun, Xudong and Schou, Jesper and Couvidat, Sebastien and Norton, Aimee and Bobra, Monica and Centeno, Rebecca and Leka, K. D. and Barnes, Graham and Turmon, Michael},
    number = {9},
    month = {9},
    pages = {3483--3530},
    volume = {289},
    publisher = {Kluwer Academic Publishers},
    url = {http://link.springer.com/10.1007/s11207-014-0516-8},
    doi = {10.1007/s11207-014-0516-8},
    issn = {0038-0938}
}

@article{Koomanski2007TheKernel,
    title = {{The interaction of a plasmoid with a loop-top kernel}},
    year = {2007},
    journal = {Astronomy {\&} Astrophysics},
    author = {Ko{\l}oma{\'{n}}ski, S. and Karlick{\'{y}}, M.},
    number = {2},
    month = {11},
    pages = {685--693},
    volume = {475},
    doi = {10.1051/0004-6361:20078356},
    issn = {0004-6361}
}

@article{DePontieu2014TheIRIS,
    title = {{The Interface Region Imaging Spectrograph (IRIS)}},
    year = {2014},
    journal = {Solar Physics},
    author = {De Pontieu, B. and Title, A. M. and Lemen, J. R. and Kushner, G. D. and Akin, D. J. and Allard, B. and Berger, T. and Boerner, P. and Cheung, M. and Chou, C. and Drake, J. F. and Duncan, D. W. and Freeland, S. and Heyman, G. F. and Hoffman, C. and Hurlburt, N. E. and Lindgren, R. W. and Mathur, D. and Rehse, R. and Sabolish, D. and Seguin, R. and Schrijver, C. J. and Tarbell, T. D. and W{\"{u}}lser, J.-P. and Wolfson, C. J. and Yanari, C. and Mudge, J. and Nguyen-Phuc, N. and Timmons, R. and van Bezooijen, R. and Weingrod, I. and Brookner, R. and Butcher, G. and Dougherty, B. and Eder, J. and Knagenhjelm, V. and Larsen, S. and Mansir, D. and Phan, L. and Boyle, P. and Cheimets, P. N. and DeLuca, E. E. and Golub, L. and Gates, R. and Hertz, E. and McKillop, S. and Park, S. and Perry, T. and Podgorski, W. A. and Reeves, K. and Saar, S. and Testa, P. and Tian, H. and Weber, M. and Dunn, C. and Eccles, S. and Jaeggli, S. A. and Kankelborg, C. C. and Mashburn, K. and Pust, N. and Springer, L. and Carvalho, R. and Kleint, L. and Marmie, J. and Mazmanian, E. and Pereira, T. M. D. and Sawyer, S. and Strong, J. and Worden, S. P. and Carlsson, M. and Hansteen, V. H. and Leenaarts, J. and Wiesmann, M. and Aloise, J. and Chu, K.-C. and Bush, R. I. and Scherrer, P. H. and Brekke, P. and Martinez-Sykora, J. and Lites, B. W. and McIntosh, S. W. and Uitenbroek, H. and Okamoto, T. J. and Gummin, M. A. and Auker, G. and Jerram, P. and Pool, P. and Waltham, N.},
    number = {7},
    month = {7},
    pages = {2733--2779},
    volume = {289},
    publisher = {Kluwer Academic Publishers},
    url = {http://link.springer.com/10.1007/s11207-014-0485-y},
    doi = {10.1007/s11207-014-0485-y},
    issn = {0038-0938}
}

@article{Priest2002TheFlares,
    title = {{The magnetic nature of solar flares}},
    year = {2002},
    journal = {The Astronomy and Astrophysics Review},
    author = {Priest, E.R. and Forbes, T.G.},
    number = {4},
    month = {3},
    pages = {313--377},
    volume = {10},
    doi = {10.1007/s001590100013},
    issn = {0935-4956}
}

@article{Fletcher2001TheRibbons,
    title = {{The Magnetic Structure and Generation of EUV Flare Ribbons}},
    year = {2001},
    journal = {Solar Physics},
    author = {Fletcher, L. and Hudson, H.},
    number = {1/2},
    pages = {69--89},
    volume = {204},
    doi = {10.1023/A:1014275821318},
    issn = {00380938}
}

@article{Karpen2012THEFLARES,
    title = {{THE MECHANISMS FOR THE ONSET AND EXPLOSIVE ERUPTION OF CORONAL MASS EJECTIONS AND ERUPTIVE FLARES}},
    year = {2012},
    journal = {The Astrophysical Journal},
    author = {Karpen, J. T. and Antiochos, S. K. and DeVore, C. R.},
    number = {1},
    month = {11},
    pages = {81},
    volume = {760},
    doi = {10.1088/0004-637X/760/1/81},
    issn = {0004-637X}
}

@article{Goode2010TheEST,
    title = {{The NST: First results and some lessons for ATST and EST}},
    year = {2010},
    journal = {Astronomische Nachrichten},
    author = {Goode, P.R. and Coulter, R. and Gorceix, N. and Yurchyshyn, V. and Cao, W.},
    number = {6},
    month = {6},
    pages = {620--623},
    volume = {331},
    doi = {10.1002/asna.201011387},
    issn = {0004-6337}
}

@article{Bradski2000TheLibrary,
    title = {{The OpenCV Library}},
    year = {2000},
    journal = {Dr. Dobb's Journal of Software Tools},
    author = {Bradski, Gary},
    number = {11},
    pages = {122--125},
    volume = {25},
    url = {https://drdobbs.com/open-source/the-opencv-library/184404319?queryText=opencv}
}

@article{Shen2022TheFlares,
    title = {{The origin of underdense plasma downflows associated with magnetic reconnection in solar flares}},
    year = {2022},
    journal = {Nature Astronomy},
    author = {Shen, Chengcai and Chen, Bin and Reeves, Katharine K. and Yu, Sijie and Polito, Vanessa and Xie, Xiaoyan},
    number = {3},
    month = {1},
    pages = {317--324},
    volume = {6},
    doi = {10.1038/s41550-021-01570-2},
    issn = {2397-3366}
}

@article{Nishizuka2009THEFLARE,
    title = {{THE POWER-LAW DISTRIBUTION OF FLARE KERNELS AND FRACTAL CURRENT SHEETS IN A SOLAR FLARE}},
    year = {2009},
    journal = {The Astrophysical Journal},
    author = {Nishizuka, N. and Asai, A. and Takasaki, H. and Kurokawa, H. and Shibata, K.},
    number = {1},
    month = {3},
    pages = {L74-L78},
    volume = {694},
    doi = {10.1088/0004-637X/694/1/L74},
    issn = {0004-637X}
}
\end{document}